\documentclass[12pt]{article}
\usepackage{epsf}
\usepackage{amsmath,amssymb}
\usepackage{amsfonts}

\newcommand{\pa}{\partial}
\newcommand{\tr}{{\rm tr}}
\newcommand{\comment}[1]{}

\newcommand{\pasl}{\pa\kern-.55em /}

\newcommand{\ksl}{k\kern-.55em /}

\newcommand{\diag}{\hbox{diag}}
\DeclareFixedFont{\xiiss}{OT1}{cmss}{m}{n}{12}
\DeclareFixedFont{\ixss}{OT1}{cmss}{m}{n}{9}
\DeclareFixedFont{\cmrnine}{OT1}{cmr}{m}{n}{9}
\newcommand{\field}[1]{\mathbb{#1}}
\newcommand{\BC}{{\field C}}\newcommand{\CC}{\BC}

\newcommand{\BZ}{{\field Z}}

\newcommand{\CCs}{\hbox{\ixss C\kern-.4emI}}
\newcommand{\ZZs}{\hbox{\ixss Z\kern-.4emZ}}

\newcommand{\CA}{{\cal A}}

\newcommand{\CM}{{\cal M}}
\newcommand{\CZ}{{\cal Z}}

\newcommand{\ZA}{\CZ\CA}

\newcommand{\G}{G}
\newcommand{\oo}{orbifold of an orbifold}
\newcommand{\BA}[1]{\widehat{\field A}_{#1}}
\newcommand{\BD}[1]{\widehat{\field D}_{#1}}
\newcommand{\BE}[1]{\widehat{\field E}_{#1}}
\newcommand{\ZZ}{{\field Z}}
\newcommand{\Bone}{{\bf 1}}
\newcommand{\Btwo}{{\bf 2}}
\newcommand{\Bthree}{{\bf 3}}
\newcommand{\Bfour}{{\bf 4}}
\newcommand{\Bsix}{{\bf 6}}

\newcommand{\myfig}[3]{\begin{figure}[ht]
\begin{center}
\leavevmode
\epsfxsize=#2cm
\epsfbox{#1}
\end{center}
\caption{#3}
\label{fig:#1}
\end{figure}}

\newcommand{\uiaddress}
{{\small\it Department of Physics, University of Illinois, Urbana, IL 61801}}
\newcommand{\email}[1]{\thanks{e-mail: \tt#1}}
\newcommand{\preprint}
{\begin{flushright}\begin{small}
ILL-(TH)-00-08\\ hep-th/0012050\\ 
\end{small}\end{flushright}       
}

\begin{document}

\begin{titlepage}
        \title{
        \preprint\vspace{1.5cm}
		D-branes on Singularities: New Quivers from Old}
		\author{
        	David Berenstein,\email{berenste@pobox.hep.uiuc.edu}\;
		Vishnu Jejjala,\email{vishnu@pobox.hep.uiuc.edu}\; and    
		Robert G. Leigh\email{rgleigh@uiuc.edu}\\ 
	\uiaddress
        \\
		}
\maketitle

\begin{abstract}
In this paper we present simplifying techniques which allow one to
compute the quiver diagrams for various D-branes at (non-Abelian)
orbifold singularities with and without discrete torsion. The main idea
behind the construction is to take the orbifold of an orbifold. Many
interesting discrete groups fit into an exact sequence $N\to G\to G/N$.
As such, the orbifold $\CM/G$ is easier to compute  as $(\CM/N)/(G/N)$
and we present graphical rules which allow fast computation given
the $\CM/N$ quiver.
\end{abstract}
\end{titlepage}

\section{Introduction}\label{sec:intro}

D-branes at singularities give rise to low energy effective field
theories of phenomenological interest.  The idea behind geometric
engineering \cite{KKV, KV, BJPSV, KMV} is to look at the gauge theories
that arise on branes at singularities.  Another approach is to study
the gauge theories that result from the intersection of branes at
angles \cite{BDL} (see, for example, \cite{bgkl, Ur1, Ur2} and references
therein).

The quiver construction of Douglas and Moore \cite{DM} provides a
diagrammatic tool to visualize the field content of supersymmetric gauge
theories on branes at orbifold singularities, $\BC^n/G$.  The technique
involves enumerating the irreducible representations of $G$ and
determining how the matter fields transform.  In many cases, this
procedure is difficult as it involves calculating the full (projective)
representation theory of some group which might be large. Here  we find
a technique whereby the quiver can be constructed without knowing the
full details of the representation theory of $G$.  Such a method should
be useful to model builders.

We present a systematic technique for obtaining quivers of low energy
gauge theories corresponding to string orbifold singularities.  
For the simplest
orbifolds, based on the Abelian discrete groups, our methods reduce to
known results.  They are most useful for orbifolds with discrete torsion
\cite{V} and for orbifolds involving non-Abelian discrete groups.  The
technique we present may also be generalized to other types of
singularities.

We consider a number of examples of supersymmetric orbifolds of $\CM=
\BC^n$ by a group $\G$, which is a discrete subgroup of $SU(n)$. $G$ is
taken to act linearly on the coordinates of $\BC^n$ in a particular
representation of $\G$.  Many of the discrete subgroups of $SU(n)$ are
(semi-) direct product groups.  Because of this, the groups fit into
exact sequences of the form
\begin{equation}
0 \rightarrow N \rightarrow \G \rightarrow \G/N \rightarrow 0 \label{eq:exact}
\end{equation}
for $N \vartriangleleft \G$ a normal subgroup of $\G$.  We are
particularly interested in studying the case where the group $G/N$ is
Abelian, as here the technique we present is most directly applicable. 
The orbifolding procedure can be thought of as the quotient
\begin{equation}\label{eq:quot}
\CM/\G = (\CM/N)/(\G/N),
\end{equation}
that is, as an \oo. Since $N,\G/N$ are smaller groups than $\G$, we
expect that the construction of the quiver diagram for $\G$ using
(\ref{eq:quot}) may be simpler. Indeed, it is well known that the
representation theory of a group $G$ may be organized in terms of the
representation theory of a normal subgroup.

Consider the Abelian case $\G= \BZ_m \times \BZ_n$, for example. There
are various possible quiver diagrams corresponding to this orbifold,
depending on the choice of discrete torsion.  It was realized that
discrete torsion acts via projective representations of the orbifold
group $G$ \cite{D, DF} (for related work, see for example \cite{BL,
BJL1, BJL2,G, Asp1}).  In the \oo\ approach, the discrete torsion is
encoded as a choice of $\BZ_n$ action on the quiver of $\CM/\BZ_m$. This
can also be seen as choices of monodromies of nodes in the quiver
$\CM/N$ under the orbifold group $G/N$ as encountered in Refs.
\cite{W2,BCD}.

Less is known about the details of how discrete torsion acts in the
non-Abelian case.  An orbifold of $\BC^3$ by the ordinary tetrahedral
group ${\field E}_6$ was recently analyzed \cite{Asp1}.  Subsequently,
the authors of Ref. \cite{amihe} calculated the discrete torsion for a
number of non-Abelian groups and examined the ordinary dihedral groups
${\field D}_k$ in detail. We extend these results by looking at several
discrete groups that fit into exact sequences with Abelian $G/N$.

One interesting fact that comes out of this analysis is that many
orbifold theories are on the same moduli space of couplings of the low
energy field theory. That is, unrelated orbifolds may give the same 
quiver diagram; the gauge theories differ only in their superpotentials.

The organization of our paper is as follows.  In Section \ref{sec:two},
we present the \oo\ construction without discrete torsion and provide a
simple diagrammatic prescription.  Section \ref{sec:introex} provides
two introductory examples.  We obtain the quivers of some non-Abelian
subgroups of $SU(2)$ by employing exact sequences of the form
(\ref{eq:exact}), and we study the quiver of $\ZZ_4$ to illustrate a
subtlety in our approach. In Section \ref{sec:prodgrp} we build in
discrete torsion. We concentrate in this section on direct product
groups $G_1 \times G_2$. In Section \ref{sec:manyex}, we consider a
number of examples which are discrete subgroups of $SU(3)$.  Section
\ref{sec:dualities} examines the different superpotentials that
correspond to the $\BA{2}$ quiver depending upon the details of how that
quiver is obtained. 

\section{The Construction without Discrete Torsion}\label{sec:two}

To begin, we disregard the possibility of discrete torsion and focus on
understanding the orbifold $\CM/\G$ in terms of the orbifold
$(\CM/N)/(\G/N)$. First, we must construct the quiver diagram of $\CM/N$
and then consider the action of the Abelian group $\G/N$ as an
automorphism of this quiver. Later we will specialize to the case $G/N$
abelian where the construction simplifies.

\newcommand{\n}{g}
We first construct the quiver diagram of $\CM/N$ using standard methods
\cite{DM}. The nodes are given by the irreducible representations of
$N$, which may be deduced from the group algebra of $N$. The underlying
vector space of the group algebra $\CA(N)$ is that of the regular
representation of $N$. If $\n_i\in N$, then the group algebra of $N$
consists of linear combinations of the form
\begin{equation}\label{eq:grpalgrel}
a = \sum a_i \n_i
\end{equation}
with $a_i\in \BC$.

Each irreducible representation determines a projector in the group
algebra which belongs to the center of the algebra, and the linear span
of these projectors generates the center of the group algebra itself.

The center of the algebra is straightforward to calculate. Indeed, for
$\sum a_i \n_i$ to commute with the generators $g$
\begin{equation}
\sum a_i g \n_i g^{-1} = \sum a_i \n_i,
\end{equation}
we need that the coefficient of $\n_i$ on both sides be the same. Thus
$a_i = a_j$ whenever there exists a $g$ such that $ g\n_i g^{-1} =
\n_j$, that is, whenever $\n_i$ and $\n_j$ belong to the same conjugacy
class $[\n_i]$ of $N$. Thus the center of the group algebra $\CZ\CA(N)$
is generated by the conjugacy classes of elements of the group $N$. The
relation between the idempotents and the conjugacy classes is a discrete
Fourier transform.

Consider now the group $\G$. As $N$ is a normal subgroup of $\G$, then
conjugation by elements of $\G$ leaves $N$ invariant. Indeed,
conjugation by an  element of $\G$ induces a group automorphism of $N$
and thus also an automorphism of the algebra $\CA(N)$ to itself. Any
automorphism of the algebra will leave the center fixed, and thus the
action of $\G$ will act as a linear transformation on $\CZ\CA(N)$. In
particular, the action will take idempotents to idempotents, so it will
permute the irreducible representations of $N$.

If $g\in \G$, then conjugation by $ g w$ with $w\in N$ will produce the
same action on the conjugacy classes of $N$ as $g$. Conjugation thus
factors on $\G/N$. Since the nodes of the quiver diagram denote
representations of $N$, $\G/N$ will act on the quiver by permuting
nodes. The action will also permute the arrows of the quiver with a
twisting by some representation of the group $\G/N$.

To understand the quiver for $\G$, we must then study the irreducible
representations of the group $\G/N$, the irreducible representations of
$N$, and the action of $\G/N$ on the quiver of $N$. Denote by $P_k$ the
projectors for the irreducible representations of $N$. We need to
understand the algebra generated by $P_k$ and the group algebra of
$\G/N$. Let $\sigma_k$ be the list of elements of $\G/N$. From the above
discussion, we have
\begin{equation}
\sigma_k P_\ell \sigma^{-1}_k = P_{\sigma_k(\ell)},
\end{equation}
which is a tensor algebra twisted by the action of $\G/N$ on $\ZA(N)$.

Any element of the algebra $\CA(\G)$ can be written in the form
\begin{equation}
\sum _{\ell k} a_{\ell,k} P_\ell \sigma_k,
\end{equation}
and we want to know which linear combinations are in the center of this 
algebra. Elements of the center must commute with all the $\sigma_m$:
\begin{equation}
\sum _{\ell k} a_{\ell,k} P_\ell \sigma_k= 
\sum _{\ell k} a_{\ell,k} P_{\sigma_m(\ell)}
 \sigma_m\sigma_k\sigma_m^{-1}.
\end{equation}

In the case where $G/N$ is Abelian, we obtain
\begin{equation}
a_{i,k} = a_{\sigma^{-1}_m(i), k}.
\end{equation}
Elements of the center must also commute with $P_\ell$. We then find
\begin{equation}
\sum_ka_{i,k}\left(P_i-P_{\sigma_k(i)}\right)\sigma_k=0.
\end{equation}
Thus elements of the center can have non-zero $a_{i,k}$ only when
$\sigma_k$ acts trivially on $P_i$. For each $P_{k}$ the group $G/N$
will generate an orbit of irreducible representations. In the above
result, we get one element of the center for each of these orbits and
for each $e \in G/N$ which leaves the $P_k$ fixed.

The orbit of $P$ is a representation of the group $G/N$, and the
elements which leave the $P$ fixed is a (normal) subgroup of $G/N$. The
projectors built out of the $e$ are in one to one correspondence with
the irreducible representations of this subgroup of $G/N$. If the orbit
of $P_k$ has $d$ elements, we get $|G/N|/d$ irreducible representations,
each of dimension $d \dim P_k$. In particular, if a $P$ is fixed under
$G/N$, the node associated to the representation splits into $|G/N|$
nodes, and if no element of $G/N$ leaves $P$ invariant, then the orbit
of $P$ contracts to a single node.

The result above determines the nodes of the new quiver diagram. We also
need to find the arrows of the diagram corresponding to the matter
fields. In a standard orbifold, the arrows are determined by 
representations of $G$ on the normal directions to the orbifold and on
the fermions. On a supersymmetric quiver these two are related, and the
arrows are determined by the action of $G$ on the directions which are
transverse to the orbifold.

In our case, the quiver of $N$ is given, and we have to understand the
action of $G/N$ on the arrows of the quiver $N$. If we permute the
nodes, we also permute the arrows between the nodes, but there might be
some extra action of $G/N$ on the arrows.

As the original action of $G$ on the variables normal to the orbifold is
linear, we expect $G/N$ to act linearly on the arrows of the quiver $N$.
If two nodes are connected in the original quiver diagram, this means
that the two are connected by a representation of $N$. $N$ is still a
subgroup of $G$, and the representations of $G$ can be split into
representations of $N$. It follows that if two orbits are connected in
$N$, the new nodes of the two orbits are connected in some way by arrows
in $G$.

Now, the orbifold action assigns an associated representation of $G/N$
to each transverse field. Given a pair of nodes corresponding to two
different orbits, we can decide if there is an arrow connecting them by
looking at the tensor product of the corresponding representations of
the subgroup of $G/N$ with the representation associated to a transverse
field.

These considerations give us the following rules.

\begin{enumerate}
\item The nodes of the new quiver diagram are obtained by splitting and
joining nodes in the old quiver diagram according to the counting given
by the orbit of the node under $G/N$. All the nodes in the orbit have
the same rank. 
\item Two orbits which are connected by arrows 
in the quiver of $N$ are connected by arrows in the quiver of $G$.
\item If the orbits have splitting nodes, then the arrows connect nodes
that are split according to the representation of $G/N$ on the arrows.
\end{enumerate}

The above rules also work if the linear action of $G/N$ closes only up
to gauge tranformations. The technical point which is distinct is that
we don't use representations of $G/N$ on the arrows, but representations
of the lift of $G/N$ in $Aut(N)$, the group of automorphisms of the
quiver.\footnote{As the arrows represent chiral fields, these are not
gauge invariant, and an action of $G/N$ on the arrows of $N$ is only
well defined up to gauge transformations. Gauge tranformations
correspond to inner automorphisms $Inn(N)$ of the quiver, and thus the
action of $G/N$ is an {\it outer} automorphism of the quiver diagram,
$Out(N) = Aut(N) /Inn(N)$.} This lift is still finite, but the group is
larger than $G/N$, and it is not true that the fields are
representations of $G/N$ in general. Some still might be, whereas some
others might not be.

\section{New Quivers from Old}\label{sec:introex}

The best way to understand the application of these rules is to 
consider a number of examples.
First, we will consider the $\BD{k}$ singularity. A seemingly
trivial second example is provided by $\BZ_4$. However, there is an
important subtlety that arises here which is important for our later
discussions.

\subsection{Example:  $\BC^2/\BD{k}$}
\newcommand{\BDo}{{\field D}}

We consider the binary dihedral group $\BD{k}$, which is the $\ZZ_2$ extension
of the ordinary dihedral group $\BDo_k$. $\BD{k}$ has no discrete torsion 
({\it i.e.} $H^2(\BD{k},U(1))=0$) and thus there is only one choice of orbifold.

Consider the exact sequence 
\begin{equation}
0 \rightarrow \ZZ_{2k} \rightarrow \BD{k} 
\rightarrow \ZZ_{2} \rightarrow 0 \label{eq:Dk-exact}.
\end{equation}
This means that $\BD{k}$ has a normal subgroup $\ZZ_{2k}$. We want to show that
the quiver for the binary dihedral group may be obtained from the $\ZZ_{2k}$
quiver in a natural way.  In other words, the exact sequence
(\ref{eq:Dk-exact}) permits us to think about the orbifold
$\BC^2/\BD{k}$ in terms of a $\ZZ_2$ action on the orbifold
$\BC^2/\ZZ_{2k}$ ({\em i.e.} $\BC^2/\BD{k} \simeq
(\BC^2/\ZZ_{2k})/\ZZ_{2}$).

The only irreducible representations of the cyclic group $\ZZ_n$ are $n$-th
roots of unity, ${\bf 1}_a = \omega_n^a$, where $\omega_n \equiv e^{2\pi
i/n}$ and the index $a = 0, 1, \ldots, n-1$.  The quiver for the orbifold
$\BC^2/\ZZ_n$ is the $\BA{n-1}$ quiver in the $A$-$D$-$E$ classification.

The binary dihedral group 
\cite{Cox-1} is 
generated by two elements satisfying
\begin{equation}
e_1 e_2 = e_2 e_1^{-1},\quad
e_1^k = e_2^2,\quad
e_1^{2k} = 1.\label{eq:Dkgens}
\end{equation}
This group, of order $4k$, has four one-dimensional irreducible
representations, which we label ${\bf 1}_j$:
\begin{eqnarray}
&{\bf 1}_0 :& (+1,+1),\label{eq:onezero}\\
&{\bf 1}_1 :& (-1,+e^{k \pi i/2}),\label{eq:oneone}\\
&{\bf 1}_2 :& (+1,-1),\label{eq:onetwo}\\
&{\bf 1}_3 :& (-1,-e^{k \pi i/2})\label{eq:onethree}
\end{eqnarray}
and $(k-1)$ two-dimensional irreducible representations, which we label
${\bf 2}_a$.  A particular choice of basis for the ${\bf 2}$s is
\begin{equation}
e_1=\begin{pmatrix}\omega_{2k}^a & 0 \\ 0 & 
\omega_{2k}^{-a}\end{pmatrix},\quad
e_2=i^a \begin{pmatrix}0 & 1 \\ 1 & 0\end{pmatrix},
\end{equation}
with the index $a = 1, \ldots, k-1$. All other representations of
$\BD{k}$ are either reducible or are $GL(2,\BC)$ equivalent to the ones
listed here.  (In particular, ${\bf 2}_{k-l}$ and ${\bf 2}_{k+l}$ are
$GL(2,\BC)$ equivalent.)  The quiver for this theory is shown in Figure
\ref{fig:dkquiver.eps}.
\myfig{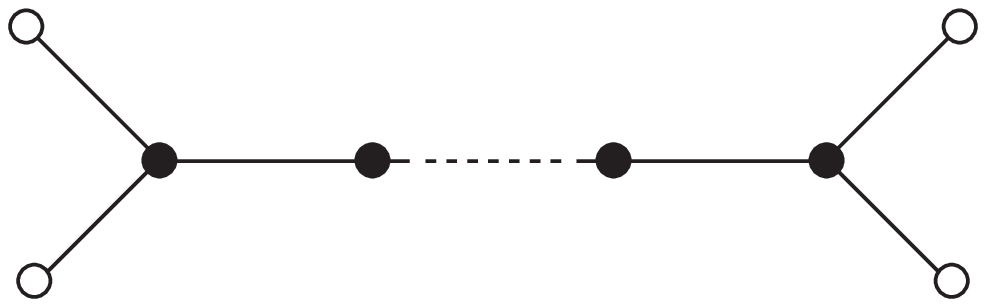}{4}{Quiver of $\BD{k}$ singularity.}
(Here we use $N=2$ notation, so a line corresponds to a hypermultiplet (a pair
of arrows).)

Now let us reproduce these results using the techniques that we have
discussed above. First we must identify the action of $\ZZ_2$ on the
$\BA{n-1}$ quiver. It is easily seen that the $\ZZ_2$ acts by
identifying a root of unity with its inverse.
\myfig{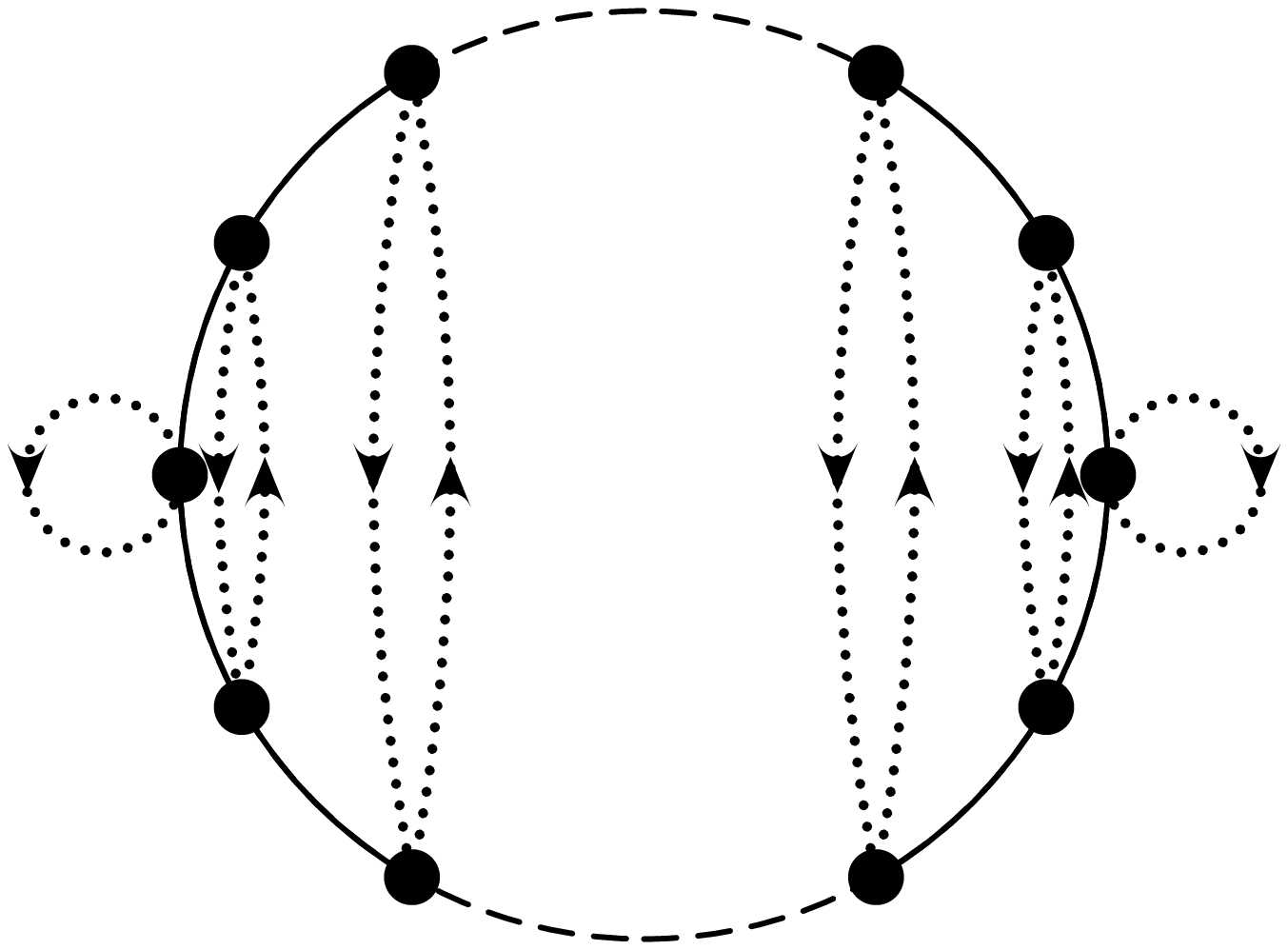}{4}{Quiver of $\BA{2k-1}$ singularity
with a $\ZZ_2$ action.} 
Thus, the node for $\omega_{2k}^{k-a}$ is identified with
$\omega_{2k}^{k+a}$ for $a = 0, 1, \ldots, k$.  Note that under this
$\ZZ_2$ action the nodes corresponding to $\omega_{2k}^0 = 1$ and
$\omega_{2k}^k = -1$ map to themselves.

The nodes which are in a 2-orbit of $\BZ_2$ combine to form a 2-node of the
new quiver; a node that is in a 1-orbit splits into two 1-nodes.  Thus, we
get a total of $k-1$ two-dimensional irreducible representations and four
one-dimensional irreducible representations.  The lines connecting nodes are
inherited from the $\BZ_{2k}$ quiver:  each pair of ${\bf 1}$s connects to
one of the ${\bf 2}$s, while the ${\bf 2}$s connect to each other along a
line. Thus we reproduce the $\BD{k}$ quiver, Figure
\ref{fig:dkquiver.eps}.

The same prescription applies for the binary polyhedral groups\footnote{The
quiver for $\BE{8}$ cannot be obtained in this way, as there is no such
useful exact sequence.}
$\BE{6}$ and $\BE{7}$ since we have the exact sequences \cite{Reid}
\begin{eqnarray}
& 0 \rightarrow \BD{2} \rightarrow \BE{6} \rightarrow \ZZ_{3} \rightarrow 0 \label{eq:E6-exact}, &\label{eq:Reid-6}\\
& 0 \rightarrow \BE{6} \rightarrow \BE{7} \rightarrow \ZZ_{2} \rightarrow 0 \label{eq:E7-exact}. &\label{eq:Reid-7}
\end{eqnarray}
Diagramatically, the quivers of these are obtained in Figures
\ref{fig:d2hatz3.eps} and \ref{fig:e6hatz2.eps}, where orbits are
denoted by the dotted lines.
\myfig{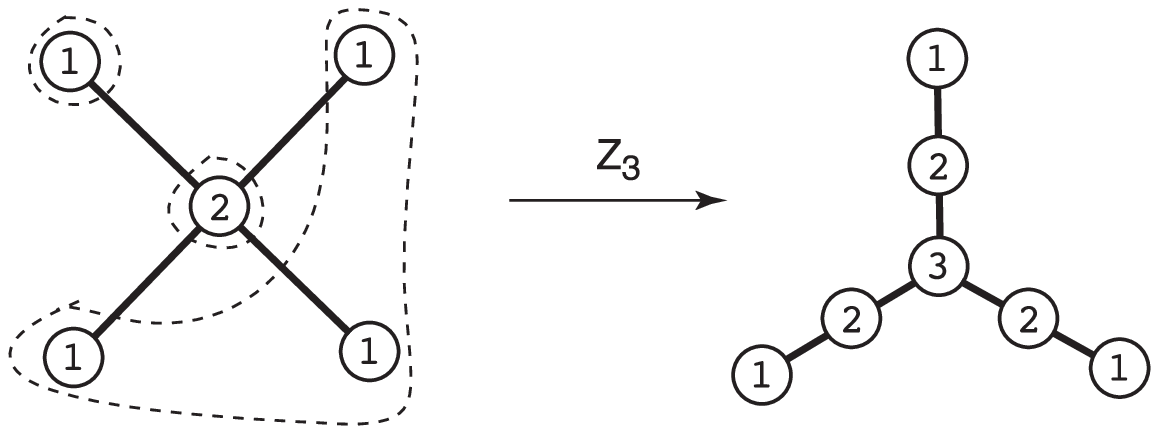}{6}{Quiver of $\BE{6}$ singularity.}
\myfig{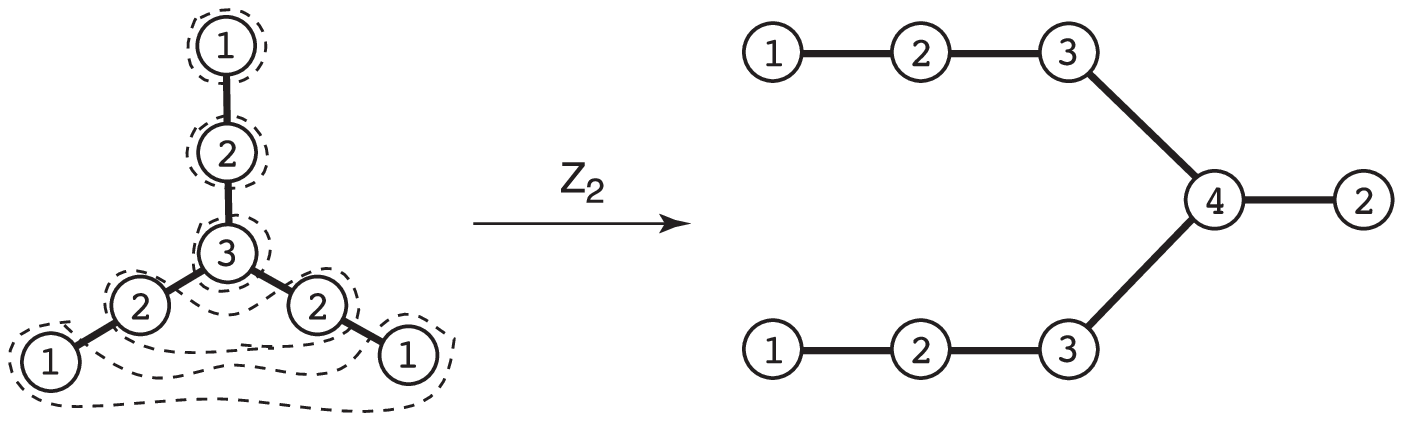}{6}{Quiver of $\BE{7}$ singularity.}

\subsection{Second Example:  $\BC^2/\ZZ_4$}

Since $\ZZ_2$ is a normal subgroup of $\ZZ_4$, we have the exact
sequence
\begin{equation} 
0 \rightarrow \ZZ_2 \rightarrow \ZZ_4 \rightarrow \ZZ_2 \rightarrow 0.
\end{equation} 
This is one of the examples where one can choose the lift of $\BZ_2$ to
the automorphisms of the quiver to involve gauge transformations. If the
left $\BZ_2$ subgroup is generated by $e$, the right $\BZ_2$ action in
the $\ZZ_2$ quiver must square to give $\sigma^2 = e$ so that we obtain
a $\BZ_4$ action. Thus we see that we must consider here an action of
the right $\ZZ_2$ which is a representation only up to a gauge
transformation.

The quiver diagram of $\ZZ_2$ consists of a pair of nodes, $+1$ and $-1$
connected by a pair of bifundamentals.  The group $\ZZ_4/\ZZ_2 \simeq
\ZZ_2$ acts trivially on $\BA{1}$ by sending each node to itself.  Rule
(1) tells us that the quiver for $\ZZ_4$ contains four nodes of rank 1. 
Rule (2) tells us that each of the nodes that formed from the splitting
of the node $+1$ connects to each of the nodes that formed from the
splitting of the node $-1$ via hypermultiplets.

Notice that in the quiver of $\BZ_2$ we have two hypermultiplets, each
corresponding to two arrows between the same two representations running
in opposite directions. We have four $N=1$ superfields, two going from
node one to node two $\phi^i_{12}$, $i=1,2$ and two going backwards
$\phi^i_{21}$. The $\BZ_2$ generated by $\sigma$ takes $\phi^1\to i
\phi^1$ and $\phi^2\to -i \phi^2$. Notice that $\sigma^2$ takes $\phi^i
\to -\phi^i$ for all $i$, and this can be interpreted as an action by
multiplying all the superfields by $-1$, which is a gauge transformation
on one of the nodes by $(-1)$.

By this construction, the nodes corresponding to $\phi^1$ transform
differently than the nodes coresponding to $\phi^2$. In particular the
arrows $\phi^1_{21}$ and $\phi^2_{21}$ will join different
representations.
Straightening out the crossed lines, we obtain the $\BA{3}$ quiver. This
is illustrated in Figure \ref{fig:babyex.eps}.
\myfig{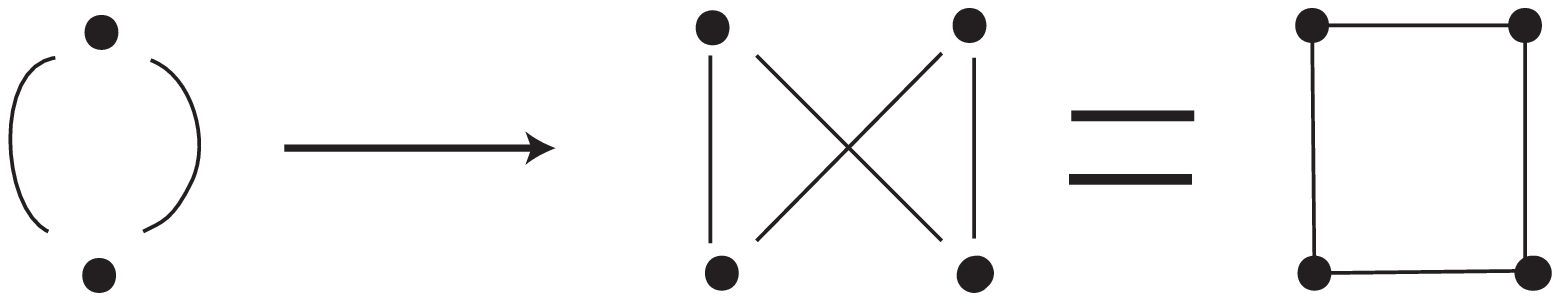}{7}{The quiver of the $\BA{3}$ singularity is 
obtained as a
$\ZZ_2$ automorphism of the quiver of the $\BA{2}$ singularity.}
The orbifold $\BC^2/\ZZ_4$ is an \oo, $(\BC^2/\ZZ_2)/\ZZ_2$.

\bigskip

Of course, these quivers may be obtained in the standard way without
difficulty. The techniques are more powerful, however, when discrete
torsion is involved.  We shall now provide a derivation of the quiver
rules listed above before analyzing these examples and others, which
include discrete torsion, in greater detail.

\section{Product Group Orbifolds}\label{sec:prodgrp}

In the general case one wants to understand how to add discrete torsion.
The purpose of this section is to address this issue. We will begin by
considering in particular the orbifold
\begin{equation}
\BC^n/ (\G_1\times \G_2),
\end{equation}
where the $\G_i$ act by linear transformations on the generators of
$\BC^n$. In later sections we consider a series of examples, generally
of non-Abelian groups with discrete torsion. All of these examples may
be thought of as (semi-) direct products, and thus it is important to
study the general case in detail.

Since the orbifold group is a direct product, we clearly have an exact
sequence
\begin{equation}
0\to\G_1\to\G_1\times \G_2\to \G_2\to 0
\end{equation}
and so, from the discussion
of the previous section, we can consider
\begin{equation}\label{eq:prodbrk}
(\BC^n/\G_1)/\G_2.
\end{equation}
The general question we want to answer now is the following:
what is the quiver diagram for this orbifold
given a choice of discrete torsion of the group $\G_1\times \G_2$?

To answer the question, first we note that there is a formula for the
discrete torsion of $\G_1\times\G_2$ \cite{Kar1}:
\begin{eqnarray}
H^2(\G_1\times \G_2,U(1))
&=& H^2(\G_1,U(1))\times H^2(\G_2,U(1))\\ &&\times 
\left[ H^1(\G_1,U(1))\otimes_\BZ H^1(\G_2,U(1))\right].\nonumber
\end{eqnarray}
Alternately, by a theorem of Yamazaki \cite{Kar1}, we may write
\begin{equation}
H^1(\G_1,U(1))\otimes_\BZ H^1(\G_2,U(1))
=Hom_\BZ (\G_1/\G_1',\G_2/\G_2'),
\end{equation}
where we have $\G'\equiv [\G,\G]$, the commutator subgroup of $G$.

Thus, discrete torsion has several sources. In light of the structure
(\ref{eq:prodbrk}), if $H^2(\G_1,U(1))\neq 0$, this should be taken into
account for the quiver corresponding to $\BC^n/\G_1$. We will
concentrate on the case where $H^2(\G_2,U(1))=0$; this is an important
simplifying assumption, and in any case is true for all of the examples
that we consider.
The remaining source of discrete
torsion is the ``interaction'' term $H^1(\G_1,U(1))\otimes_\BZ
H^1(\G_2,U(1))$. The group $H^1(\G,U(1))$ is the group of
one-dimensional irreducible representations of $G$, with multiplication
given by the tensor product of representations. 
Generally, we will refer to elements of this group by $\chi$.

\subsection{Abelian $\G_2$}

The case of Abelian $G_2$ is particularly
simple. If $\G_2$ is the cyclic group $\BZ_n$, we have
 $H^1(\G_2,U(1))= \G_2 $, the group of characters of 
$\G_2$.
In this case we have
\begin{equation}
H^1(\G_1,U(1))\otimes_\BZ \BZ_n \subseteq H^1(\G_1,U(1)).
\end{equation}
A choice of discrete torsion amounts to a choice of subgroup of the
characters of $G_1$. 
We now need to consider representations of the group algebra of 
$\G_1\times \BZ_n$ with the choice of discrete torsion. 

We will write $e$ as the generator of $\G_2$. The group algebra
$\CA(\G)$ will be generated by $e$ and the generators of $\G_1$. Because
$\G$ is a product group, $g e$ and $eg$ can differ only by a phase, and
in fact
\begin{equation}\label{eq:distor}
e\cdot g= g\cdot e \chi(g)
\end{equation}
for any $g\in \G_1$ and $\chi\in H^1(\G_1,U(1))$. As a result, we can
think of $e$ as acting on the elements (\ref{eq:grpalgrel}) of the group
algebra
\begin{equation}
e:\sum_i a_i g_i \to e\cdot\sum_i a_i g_i\cdot e^{-1},
\end{equation}
and this is an outer automorphism of
$\CA(\G_1)$. In particular, the action of $e$ leaves the center
$\ZA(\G_1)$ invariant. We conclude that it acts on the representations
of $\G_1$ by permutations. Because of (\ref{eq:distor}), $e$ takes a
representation $R$ to $\chi\otimes R$, which is irreducible and has the
same dimension as $R$.

A first step then in constructing the quiver of $(\CC^n/\G_1)/\G_2$ is
to decide on how $e$ permutes the nodes of the $\CC^n/\G_1$ quiver. The
bottom line is that different choices of discrete torsion correspond
to different sets of orbits of nodes. 

We will discuss the representations of $\G_1$ in terms of projectors
$P_i$, as in Section \ref{sec:two}. In the present case, every element
of the center of the algebra can be written as
\begin{equation}
\sum_{i,\ell} a_{i\ell} P_i e^\ell,
\end{equation}
where  $a_{j\ell} = a_{\sigma^{-1}_e(j),\ell}$. 

We also have
\begin{equation}
\sum_\ell a_{j\ell} (P_j-P_{\sigma_\ell(j)}) e^l=0
\end{equation}
for all $j$. That is, only if $e^k$ acts trivially on the orbit of
$P_j$, can we have $a_{jk} \ne 0$. Thus for each orbit of
representations of the group $\G_1$ of order $k'$, we get $n/k'$
distinct representations, which appear as nodes in the new quiver. Note
that a given value of $k$ is a multiple of $k'$.

To be more specific, we implement the action of $e$, Eq.
(\ref{eq:distor}). Choose a basis $|v_i\rangle_R$ for the vector space
where $\G_1$ acts block diagonally. Then $e|v\rangle_R$ transforms under
$\G_1$ as $|v\rangle_{\chi\otimes R}$. Consider a matrix representation
for $e$; we can write this in terms of an invertible matrix $U$ as
\begin{equation}
e |v_i\rangle_R = (U_i^k)_{R, \chi\otimes R}|v_k\rangle_{\chi \otimes R}.
\end{equation}
If we think of elements of $\G_1$ as block diagonal
\begin{equation}
R(\G_1) = \diag( R_0, R_1, R_2, \dots, R_{k'-1}),
\end{equation} 
then $e$ takes the form
\begin{equation}
e \sim \begin{pmatrix} 0 & U_{R_0, R_1} &0 &\dots&0\\
0 & 0 & U_{R_1, R_2} &\dots&0\\
\vdots & \vdots & \vdots &\ddots &\vdots\\
U_{R_{k'-1},R_{0}}& 0&0&\dots &0 
\end{pmatrix}.
\end{equation}
Here, we have a set of $k'$ $d$-dimensional representations $ R_0, R_1,
R_2, \dots, R_{k'-1}$ that are permuted by $e$ (that is,
$R_i=\chi\otimes R_{i-1}$). By $GL(k'd,\CC)$ transformations, we can set
the determinants of each submatrix $U_{R_i,R_j}$ to unity, leaving an
overall phase $\alpha$ in $e$. Since $e^n=I$, we must have $\alpha^n=1$.
By acting with $e^k$, we get an element which belongs to the center of
the twisted group algebra. On an irreducible represention this is
proportional to the identity. Thus $e^k = \diag((U_{R_0,R_1}\dots
U_{R_{k-1} R_0})^{k/k'}\alpha^ k) $. Without loss of generality, we can
take $(U_{R_0,R_1}\dots U_{R_{k-1} R_0})^{k/k'} = I$, and thus $e^k =
\alpha^k I$.

Since $\alpha$ is an $n$-th root of unity it might appear that we get
$n$ distinct representations this way. However, there is an $SL(k'd)$
transformation that takes $\alpha\to\alpha\omega_{k'}$, where
$\omega_{k'}^{k'}=1$. Thus, we reproduce the result that there are
$n/k'$ distinct irreducible representations of dimension $dk'$ of the
group for each of these orbits.

Notice that if we sum the square of the dimensions of the
representations over all orbits  we get
\begin{equation}
\sum_{orb} \frac{n}{k'} (k'd)^2 = \sum_{orb} n d^2 k'= n \sum_{orb} d^2 k'.
\end{equation}
Each orbit is made of exactly $k'$ distinct irreducible (projective)
representations of $\G_1$, so the sum over orbits covers the sum over
irreducible representations of $\G_1$. Thus
\begin{equation}
\sum_{orb} \frac{n}{k'} (k'd)^2 = n \sum_{R} \dim(R)^2 = n |\G_1| = 
|\G_1\times \BZ_n|.
\end{equation}
This equality can only hold if the sum in the left is over all the
possible distinct irreducible representations of the group $|\G_1\times
\BZ_n|$ with a given cocyle. Thus we have obtained all of the nodes of
the quiver in this way.

\subsection{Non-Abelian $G_2$}

If $\G_2$ is non-Abelian it is not true in general that $H^1(\G_2, U(1))$ is equivalent
to $\G_2$; there is an exact sequence of groups
\begin{equation}
0\to \tilde\G_2 \to \G_2\to H^1(\G_2,U(1))\to 0
\end{equation}
with $\tilde\G_2$ the kernel of the group map. In such a case, an
orbifold $\CM/\G_2$ may be understood as
$(\CM/\tilde\G_2)/H^1(\G_2,U(1))$; this reduces the non-Abelian quotient
to an Abelian one, provided we understand how the discrete torsion
behaves for a sequence of groups as above. In general this particular
problem is complicated,\footnote{The calculation of the group cohomology
groups can be carried out in their classifying spaces. Each sequence of
groups induces a fibration and one can compute some approximation to the
cohomology via a spectral sequence \cite{Adem}. The difficult point is
to know if the spectral sequence approximation and the cohomology
actually agree.} but we can apply the same ideas as before to build
representations.

\section{Examples: Discrete Subgroups of $SU(3)$}\label{sec:manyex}

Discrete torsion in orbifolds of the Abelian product groups has been
extensively studied \cite{VW, D, DF, MukRay, BL, BJL1, BJL2}, so we utilize
the tools we have developed for constructing quivers to analyze several
non-Abelian (semi-) direct product groups. The dedicated reader may find it 
instructive to apply our methods to the $\ZZ_m\times\ZZ_n$ case.
Here, we will focus on some of the non-Abelian
discrete subgroups of $SU(3)$. These have been listed recently in Ref.
\cite{amihe}, where the discrete torsion was computed. We should note here
that there is one important subtlety if we want to consider subgroups of
$SU(3)$: namely, the group that was written in Ref. \cite{amihe} as $\G\times
\BZ_n$ (where $\G=\BD{k},\BE{k}$) is in actuality $(\G\times\BZ_{2n})/\BZ_2$.
This point will be explained in the following; the results for discrete
torsion are modified accordingly. 

\subsection{$\BC^3/[(\BD{k} \times \ZZ_{2n})/\ZZ_2]$}

Since $H^2(\BD{k}, U(1))$ and $H^2(\ZZ_n, U(1))$ are both trivial, the
direct product group $\BD{k} \times \ZZ_n$ has discrete torsion given by
$Hom_\ZZ(\BD{k}/\BD{k}', \ZZ_n)$.  For even $k$, $\BD{k}/\BD{k}'$ is
$\ZZ_2 \oplus \ZZ_2$, and for odd $k$ it is $\ZZ_4$.  Thus \cite{amihe},
\begin{equation}
H^2(\BD{k}\times\ZZ_n,U(1)) = 
\left\{
\begin{matrix}
                        \ZZ_4 && k\ {\rm odd},\ n=0\ {\rm mod}\ 4,\\
                        \ZZ_2 && k\ {\rm odd},\ n=2\ {\rm mod}\ 4,\\
                        \ZZ_2\times\ZZ_2 && k\ {\rm even},\ n\ {\rm even},\\
                        1 && n\ {\rm odd}.
\end{matrix}\right. \label{eq:wrongdt}
\end{equation}
However, $\BD{k} \times \ZZ_n \subset SU(2) \times U(1)$ is not a
subgroup of $SU(3)$ for even $n$.  Rather, we should consider the group
$(\BD{k} \times \ZZ_{2n})/\ZZ_2 \subset (SU(2) \times U(1))/\ZZ_2$.

To unravel this technical point, let us consider the group action of the
generators of the product group $\BD{k} \times \ZZ_n$.  Let $e_0$ be the
generator of $\ZZ_n$ and $e_1$, $e_2$ be the generators of $\BD{k}$,
whose group algebra is given in Eq. (\ref{eq:Dkgens}).  Then the
generators act on the coordinates of $\BC^3$ as follows.
\begin{eqnarray}
e_0: && (z_1, z_2, z_3) \rightarrow (\omega_n z_1, \omega_n z_2, \omega_n^{-2} z_3), \nonumber \\
e_1: && (z_1, z_2, z_3) \rightarrow (\omega_{2k} z_1, \omega_{2k}^{-1} z_2, z_3), \label{eq:act1} \\
e_2: && (z_1, z_2, z_3) \rightarrow (i z_2, i z_1, z_3). \nonumber
\end{eqnarray}
When $n$ is even, $e_0^{n/2}$ and $e_1^k$ have the same action on the
space. The group action is not faithful.  (This was pointed out in the
context of brane box models in Refs. \cite{FHH-1, FHH-2}.)  Na\"{\i}vely
applying the quiver rules with discrete torsion given by Eq.
(\ref{eq:wrongdt}) yields incorrect results.

The group $\G = (\BD{k} \times \ZZ_{2n})/\ZZ_2$ mods out by the
ambiguity in the group action.  To determine the discrete torsion that
$\G$ admits, we examine the twisted group algebra explicitly.  We have
the relations
\begin{eqnarray}
e_0^n = e_1^k = e_2^2, & e_0^{2n} = 1, & \nonumber \\ 
e_0 e_1 = \theta e_1 e_0, & e_0 e_2 = \eta e_2 e_0, & e_1 e_2 = e_2 e_1^{-1},
\end{eqnarray}
where $\theta$ and $\eta$ are phases inherent to the projective
representations of the algebra \cite{D, DF}.  They encode the characters
of $(\BD{k} \times \ZZ_{2n})/\ZZ_2$.  The above relations imply that
\begin{equation}
\theta^n = \theta^k = \theta^2 = \eta^n = \eta^2 = 1.
\end{equation}
Thus, 
\begin{equation} 
(\theta, \eta) = 
\left\{
\begin{matrix}
(\pm 1, \pm 1) && k\ {\rm even},\ n\ {\rm even}, \\ 
(+1, \pm 1) && k\ {\rm odd},\ n\ {\rm even}, \\
(+1, +1) && n\ {\rm odd}, 
\end{matrix} \right.
\end{equation}
which means that
\begin{equation}
H^2((\BD{k}\times\ZZ_{2n})/\ZZ_2,U(1)) =
\left\{
\begin{matrix}
                        \ZZ_2 && k\ {\rm odd},\ n\ {\rm even},\\
                        \ZZ_2\times\ZZ_2 && k\ {\rm even},\ n\ {\rm even},\\
                        1 && n\ {\rm odd}.
\end{matrix}\right. \label{eq:rightdt}
\end{equation}
This result is consistent with the long exact sequence in cohomology
obtained from
\begin{equation}\label{eq:dzseq}
0 \rightarrow \ZZ_2 \rightarrow \BD{k} \times \ZZ_{2n} \rightarrow (\BD{k} \times \ZZ_{2n})/\ZZ_2 \rightarrow 0
\end{equation}
using theorems of Hochschild and Serre and Iwahori and Matsumoto, which
may be found in Ref. \cite{Kar1}.  We note that Eq. (\ref{eq:rightdt})
differs from Eq. (\ref{eq:wrongdt}) in a crucial way.  We now conclude
that whenever $k$ is odd and $n$ is even the discrete torsion is
$\ZZ_2$.  The $n = 0\ {\rm mod}\ 4$ and $n = 2\ {\rm mod}\ 4$ cases are
not distinguished.

\subsubsection{The $N=1$ Quiver} \label{sec:quiver}

%
To construct the nodes of the quiver, we can consider the sequence
\begin{equation}
0\to\BD{k}\to (\BD{k}\times \ZZ_{2n})/\ZZ_2\to\ZZ_n\to 0.
\end{equation}
The nodes of the quiver can thus be constructed by considering
a $\ZZ_n$ action on the $\BD{k}$ quiver. Without discrete torsion, all orbits
are one-dimensional, and thus we obtain a stack of $n$ copies of
the $\BD{k}$ quiver, with arrows to be determined.

We can see this structure directly.
Without discrete torsion, we have the relations
\begin{eqnarray}
e_0^n = e_1^k = e_2^2, & e_0^{2n} = 1, & \nonumber \\ 
e_0 e_1 = e_1 e_0, & e_0 e_2 = e_2 e_0, & e_1 e_2 = e_2 e_1^{-1}. \label{eq:newalg}
\end{eqnarray}
This algebra possesses $4n$ one-dimensional representations:
\begin{equation}
(e_0, e_1, e_2)  =  
\left\{ \begin{matrix}
        (\omega_{2n}^{2a}, +1, \pm 1), && \cr 
        (\omega_{2n}^{2a}, -1, \pm 1) && k\ {\rm even}, \cr
        (\omega_{2n}^{2a+1}, -1, \pm i) && k\ {\rm odd},
        \end{matrix}\right.
\end{equation}
where $a = 0, 1, \ldots, n-1$. The $n(k-1)$ two-dimensional irreducible
representations can be written as
\begin{equation}
e_0 = \begin{pmatrix}\omega_{2n}^b & 0 \\ 0 & \omega_{2n}^b\end{pmatrix},\ \ 
e_1 = \begin{pmatrix}\omega_{2k}^c & 0 \\ 0 & \omega_{2k}^{-c}\end{pmatrix},\ \ 
e_2 = \begin{pmatrix}0 & i^c \\ i^c & 0\end{pmatrix},
\end{equation}
where $b = 0, 1, \ldots {2n-1}$, $c = 1, 2, \ldots, k-1$, and $b = c\
{\rm mod}\ 2$.  The nodes of the $(\BD{k} \times \ZZ_{2n})/\ZZ_2$ quiver
are arrayed as $2n$ copies of half a $\BD{k}$ quiver.  We make the
observation that when $e_0$ is an even power of $\omega_{2n}$, $e_1$ and
$e_2$ are representations of ${\field D}_k$, whereas when the power of
$\omega_{2n}$ is odd, $e_1$ and $e_2$ are representations of ${\field
D}_k$ with $\ZZ_2$ discrete torsion turned on.

To determine how the arrows are drawn, we note the action of the
generators on $\BC^3$.
\begin{eqnarray}
e_0: && (z_1, z_2, z_3) \rightarrow (\omega_{2n} z_1, \omega_{2n} z_2, \omega_{2n}^{-2} z_3), \nonumber \\
e_1: && (z_1, z_2, z_3) \rightarrow (\omega_{2k} z_1, \omega_{2k}^{-1} z_2, z_3), \label{eq:act2} \\
e_2: && (z_1, z_2, z_3) \rightarrow (i z_2, i z_1, z_3). \nonumber
\end{eqnarray}
Thus, the orbifold group acts as $(\omega_{2n}, \Btwo_1)$ on $(\phi_1,
\phi_2)$ and as $(\omega_{2n}^{-2}, \Bone_0)$ on $\phi_3$.  To fill in
the lines of the quiver, we consider the decompositions $R\otimes
R_i=\oplus R_j$, drawing a (chiral) line from $R_i$ to $R_j$.

For ($\phi_1,\phi_2)$ we have
\begin{eqnarray}
(\omega_{2n}, \Btwo_1) \otimes (\omega_{2n}^{\ell}, \Bone_j) & = & (\omega_{2n}^{\ell+1}, \Btwo_{x(j)}),\\
(\omega_{2n}, \Btwo_1) \otimes (\omega_{2n}^{\ell}, \Btwo_a) & = & (\omega_{2n}^{\ell+1}, \Bfour_a),
\end{eqnarray}
where $x(0)=x(2)=1$ and $x(1)=x(3)=k-1$. The representations ${\bf 4_a}$
are reducible as follows: ${\bf 4}_1 = \Btwo_2 \oplus \Bone_0 \oplus
\Bone_2$, ${\bf 4}_{k-1} = \Btwo_{k-2} \oplus \Bone_1 \oplus \Bone_3$
and ${\bf 4}_a = \Btwo_{a-1} \oplus \Btwo_{a+1}$ for $a \neq 1,k-1$.

For $\phi_3$, we have
\begin{eqnarray}
(\omega_{2n}^{-2}, \Bone_0) \otimes (\omega_{2n}^{\ell}, \Bone_j) & = & (\omega_{2n}^{\ell-2}, \Bone_j),\\
(\omega_{2n}^{-2}, \Bone_0) \otimes (\omega_{2n}^{\ell}, \Btwo_a) & = & (\omega_{2n}^{\ell-2}, \Btwo_a).
\end{eqnarray}

As anticipated, the resulting quiver may be thought of as a stack of quivers of the form
of Figure \ref{fig:dkquiver.eps} with $n$ levels. The fields $\phi_1,
\phi_2, \phi_3$ cause interconnections between the levels.  This is
rather complicated to draw in general, and we will show only the
interconnections between two levels of the stack.  The $\phi_1$ and
$\phi_2$ arrows from $\Bone_2$ and $\Bone_3$ for odd $k$ are slightly
different than what is shown in Figure \ref{fig:dkzn.eps} in that they
connect the two levels of the stack.
\myfig{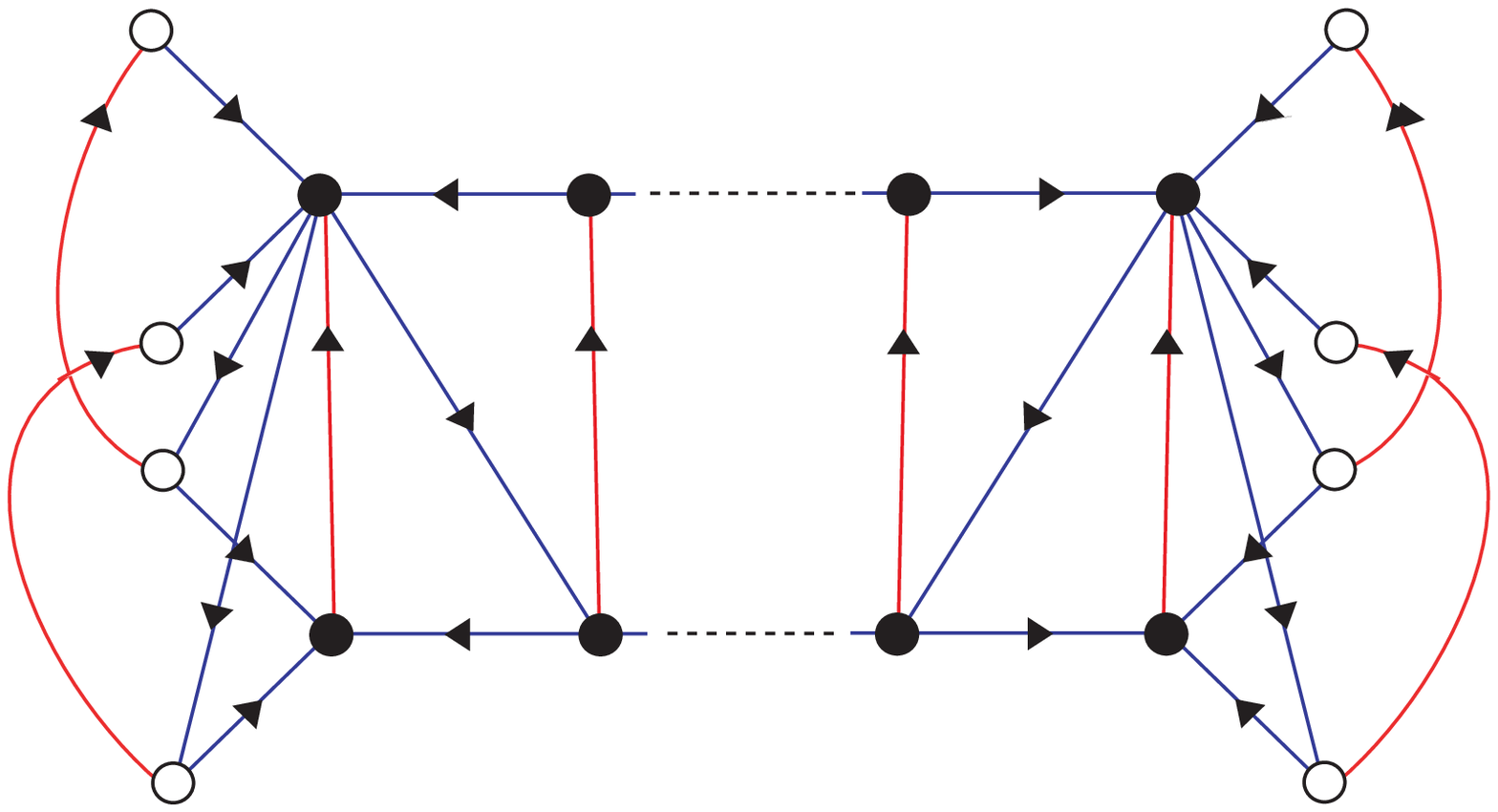}{6}{A part of the $(\BD{k} \times \ZZ_{2n})/\ZZ_2$ quiver.}

\subsubsection{$H^2((\BD{k}\times\ZZ_{2n})/\ZZ_2,U(1)) = \ZZ_2$}

When $n$ is even and $k$ is odd, the $(\BD{k}\times\ZZ_{2n})/\ZZ_2$
algebra admits a $\ZZ_2$ discrete torsion.  This is implemented as a
modification of the algebra in Eq. (\ref{eq:newalg}).  Now,
\begin{equation}
e_0 e_2 = - e_2 e_0.
\end{equation}
The only irreducible representations of this algebra are two-dimensional.  Up
to conjugation by elements of $GL(2,\BC)$, we have
\begin{equation}
e_0 = \begin{pmatrix}\omega_{2n}^a & 0 \\ 0 & -\omega_{2n}^a\end{pmatrix},\ \
e_1 = \begin{pmatrix}\omega_{2k}^b & 0 \\ 0 & \omega_{2k}^{-b}\end{pmatrix},\ \
e_2 = \begin{pmatrix}0 & i^b \\ i^b & 0\end{pmatrix}, \label{eq:dkz2twos}
\end{equation}
where $a = 0, 1, \ldots, n-1$, $b = 0, 1, \ldots, 2k-1$, and $a = b\
{\rm mod}\ 2$.  Thus, there are $nk$ $\Btwo$s.  The representations for
which $b = 0, k$ correspond to the $\Btwo$s built from combining
$\Bone$s.  All the representations found in the original quiver are
accounted for:
\begin{equation}
4n \cdot 1^2 + n(k-1) \cdot 2^2 = nk \cdot 2^2.
\end{equation} 

This is precisely what the quiver rules listed in Section
\ref{sec:introex} tell us.  Consider as an example the quiver of
$(\BD{3}\times\ZZ_4)/\ZZ_2$. Without discrete torsion, the quiver
consists of two copies of the $\BD{3}$ quiver with interconnections. 
With discrete torsion, the $\ZZ_2$ acts to combine the $\Bone_0$
and $\Bone_2$ into a $\Btwo$.  The $\Bone_1$ and $\Bone_3$
representations combine to give a second $\Btwo$.  The two $\Btwo$s in
the center of the quiver split into four $\Btwo$s since they are acted
upon trivially by $\ZZ_2$.  Thus, the final quiver consists of six
nodes, all of which are $\Btwo$s.
\myfig{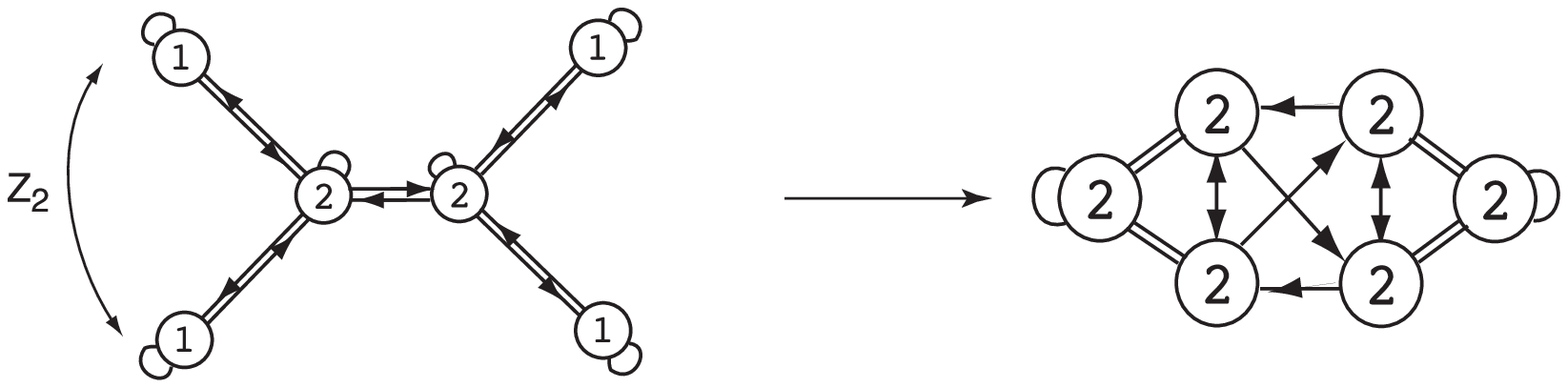}{6}{$(\BD{3} \times \ZZ_4)/\ZZ_2$ with $\ZZ_2$ torsion.}

The $\phi_3$ 
lines are adjoints for the $\Btwo$s that were constructed from the
one-dimensional irreducible representations of $\BD{3}$.  In the
$\BD{3}$ quiver, a hypermultiplet connected a $\Bone$ to its neighboring
$\Btwo$, so now, a hypermultiplet connects the $\Btwo$ built from the
$\Bone$s to each of the daughters of the neighbor which split.  The
lines between the $\Btwo$s that formed through splitting are chiral in
the new quiver, which mimics the structure of the links in the
$(\BD{3}\times\ZZ_4)/\ZZ_2$ model without discrete torsion.

The interconnections become increasingly complicated as $n$ and $k$
increase.  For example, the quiver with discrete torsion for a
$(\BD{3}\times\ZZ_{2n})/\ZZ_2$ model consists of $n/2$ interconnected copies
of the quiver shown in Figure \ref{fig:d3hatz2.eps}.  These theories are
generically chiral.

\subsubsection{$H^2((\BD{k}\times\ZZ_{2n})/\ZZ_2,U(1)) = \ZZ_2 \times \ZZ_2$}

We have $\ZZ_2 \times \ZZ_2$ discrete torsion if both $n$ and $k$ are even.
When considering the algebra, the relations in Eq. (\ref{eq:newalg}) are
modified as follows:
\begin{eqnarray}
e_0 e_1 = \pm e_1 e_0, & & e_0 e_2 = \pm e_2 e_0 \label{eq:Z2Z2alg}.
\end{eqnarray}
The four sign choices correspond to the four $\Bone$s of $\BD{k}$.  A
non-trivial automorphism of the quiver maps $\Bone_0$ to
one of the other one-dimensional irreducible representations, which then
maps back to $\Bone_0$.  

If $\Bone_0$ and $\Bone_2$ lie within an orbit, then the torsion acts
analogously to the even $n$, odd $k$ case discussed above.  We choose signs
$(+,-)$ in Eq. (\ref{eq:Z2Z2alg}).  The only irreducible representations
consistent with this choice are $\Btwo$s, and, up to $GL(2,\BC)$ equivalence,
these have the explicit realization given in Eq. (\ref{eq:dkz2twos}).  Once
again, $b = 0, k$ correspond to two-dimensional representations that form
when one-dimensional representations combine while the other values of $b$
correspond to the splitting of the two-dimensional nodes in the torsion-free
quiver.  The quiver for $(\BD{4} \times \ZZ_4)/\ZZ_2$ with this choice of
torsion is given in Figure \ref{fig:d4hatz2.eps}.
\myfig{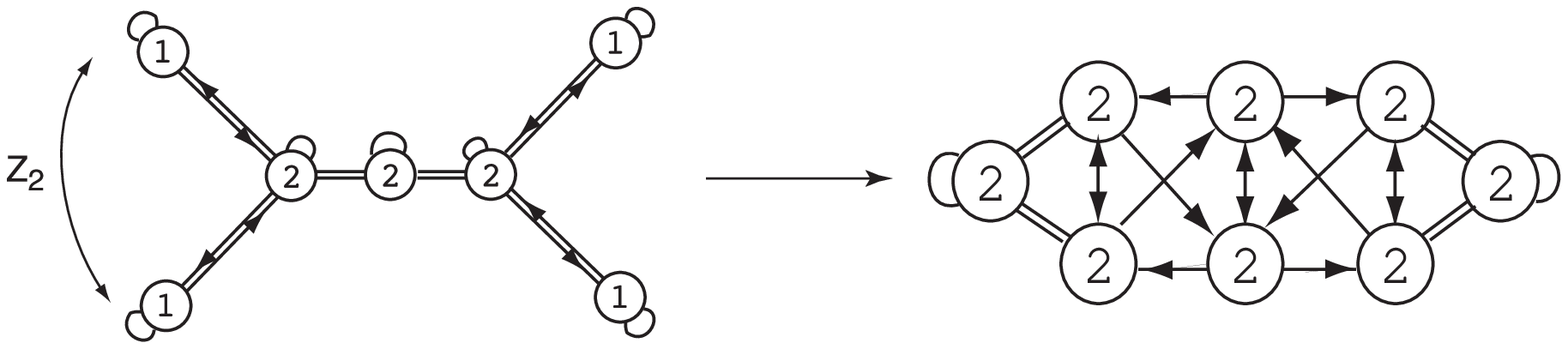}{6}{$(\BD{4} \times \ZZ_4)/\ZZ_2$ with $\ZZ_2 \times \ZZ_2$ discrete torsion:
$\Bone_0 \leftrightarrow \Bone_2$.}
In the generic case where $n > 2$, we obtain $n/2$ interconnected copies of
this quiver.

If $\Bone_0$ lies in an orbit with either $\Bone_1$ or $\Bone_3$, there
are four-dimensional irreducible representations of the algebra as well.
 The key observation here is that it is only when the discrete torsion
acts to produce orbits which cross the vertical axis of the quiver that
we get higher dimensional irreducible representations from orbits that
map pairs of $\Btwo$s to each other.  Such a result was impossible for
odd $k$ because the $\ZZ_2$ discrete torsion is incompatible with the
choice of characters corresponding to $\Bone_1$ and $\Bone_3$.  That is
to say, we could not introduce factors of $\pm i$ into the twisted
algebra.

An explicit realization of the $\Bone_0 \leftrightarrow \Bone_3$ orbit,
which corresponds to the sign choice $(-,-)$ in Eq. (\ref{eq:Z2Z2alg}),
is provided below.  The $2n$ $\Btwo$s are
\begin{equation}
e_0 = \omega_n^a \sigma^1,\;\;\; e_1 = \sigma^3,\;\;\; e_2 = \pm \sigma^3,
\end{equation}
where $\sigma^i$ are the spin-$1/2$ Pauli matrices and $a = 0, 1,
\ldots, n-1$.  The $\Bfour$s are built by using direct sums of the
representations of $\BD{k}$ for $e_1$ and $e_2$ and then solving for
$e_0$ using the relations of the twisted algebra.  An explicit
realization is
\begin{equation}
e_0 = \begin{pmatrix}0 & 0 & 0 & \alpha \\ 0 & 0 & 1 & 0 \\ 
                     0 & \alpha \omega_n^a & 0 & 0 \\ \omega_n^a & 0 & 0 & 0 \end{pmatrix},\ \ 
e_1 = \begin{pmatrix}\omega_{2k}^b & 0 & 0 & 0 \\ 0 & \omega_{2k}^{-b} & 0 & 0 \\
                     0 & 0 & \omega_{2k}^{k-b} & 0 \\ 0 & 0 & 0 & \omega_{2k}^{-(k-b)} \end{pmatrix},\ \ 
e_2 = \begin{pmatrix}0 & i^b & 0 & 0 \\ i^b & 0 & 0 & 0 \\
                     0 & 0 & 0 & i^{k-b} \\ 0 & 0 & i^{k-b} & 0 \end{pmatrix},
\end{equation}
where $a = 0, 1, \ldots, n-1$, $b = 1, 2, \ldots, (k-2)/2$, and $\alpha
\equiv (-1)^{k/2 - b + 1}$.  The $b$ compatible with a given $a$ are
those for which the relation
\begin{equation} 
(-1)^a = (-1)^b \alpha^{n/2}
\end{equation}
is true. In the end, there are $n(k-2)/4$ such $\Bfour$s.  The quiver
for the $(\BD{4} \times \ZZ_4)/\ZZ_2$ model with this choice of torsion
is given in Figure \ref{fig:d4hatz2a.eps}.
\myfig{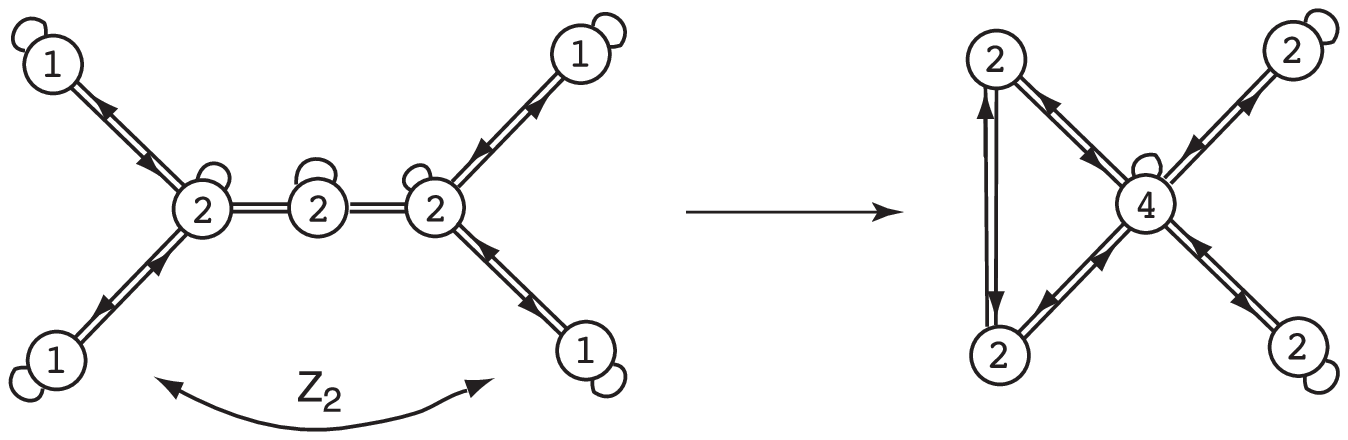}{6}{$(\BD{4} \times \ZZ_4)/\ZZ_2$ with $\ZZ_2 \times \ZZ_2$ discrete torsion:
$\Bone_0 \leftrightarrow \Bone_3$.}
Unlike the quiver obtained from the $\Bone_2$ choice of characters, this
theory is non-chiral.

\subsection{$\BC^3/[(\BE{6} \times \ZZ_{2n})/\ZZ_2]$, 
            $\BC^3/[(\BE{7} \times \ZZ_{2n})/\ZZ_2]$}

We now explore a network of exact sequences that correlate the ordinary
tetrahedral and octahedral groups to their double covers.  Algebraic details
regarding the representation theory of polyhedral groups are relegated to
Appendix ${\field E}$.  We then explore introducing discrete torsion into
orbifolds of product groups involving $\BE{6}$ and $\BE{7}$.

\subsubsection{$\BC^2/\BE{6}$ Revisited}

In Section \ref{sec:introex}, we observed that the quiver for $\BE{6}$
arises from a $\ZZ_3$ action on the quiver for $\BD{2}$ and that the
quiver for $\BE{7}$ arises from a $\ZZ_2$ action on the quiver for
$\BE{6}$ because of the exact sequences Eqs.
(\ref{eq:Reid-6}---\ref{eq:Reid-7}).  In fact, the structure is much
richer.  In the following discussion, we specialize to the case of
$\BE{6}$.  A similar story holds for $\BE{7}$.

The binary tetrahedral group is at the center of the following web of
exact sequences:
\begin{equation}
\begin{matrix}
\ZZ_2        &             & \ZZ_2        &             &       \cr
\downarrow   &             & \downarrow   &             &       \cr
\BD{2}       & \rightarrow & \BE{6}       & \rightarrow & \ZZ_3 \cr
\downarrow   &             & \downarrow   &             &       \cr             
{\field D}_2 & \rightarrow & {\field E}_6 & \rightarrow & \ZZ_3 \cr
\end{matrix}\label{eq:web}
\end{equation}
The middle horizontal line of this web is the construction of $\BE{6}$
in Section \ref{sec:introex}.  The bottom line is the analogous
construction of ${\field E}_6$ from ${\field D}_2$.  The ordinary
dihedral group ${\field D}_2$ is isomorphic to $\ZZ_2 \times \ZZ_2$,
which admits $\ZZ_2$ discrete torsion \cite{D, DF, BL}.  The ordinary
tetrahedral group has the same discrete torsion:  $H^2({\field E}_6,
U(1)) = \ZZ_2$ \cite{Kar2}. Figure \ref{fig:orde6.eps} shows the
non-chiral ${\field E}_6$ quivers, both without and with $\ZZ_2$
torsion, that are built from imposing a $\ZZ_3$ action on the ${\field
D}_2$ quiver without and with $\ZZ_2$ torsion.
\myfig{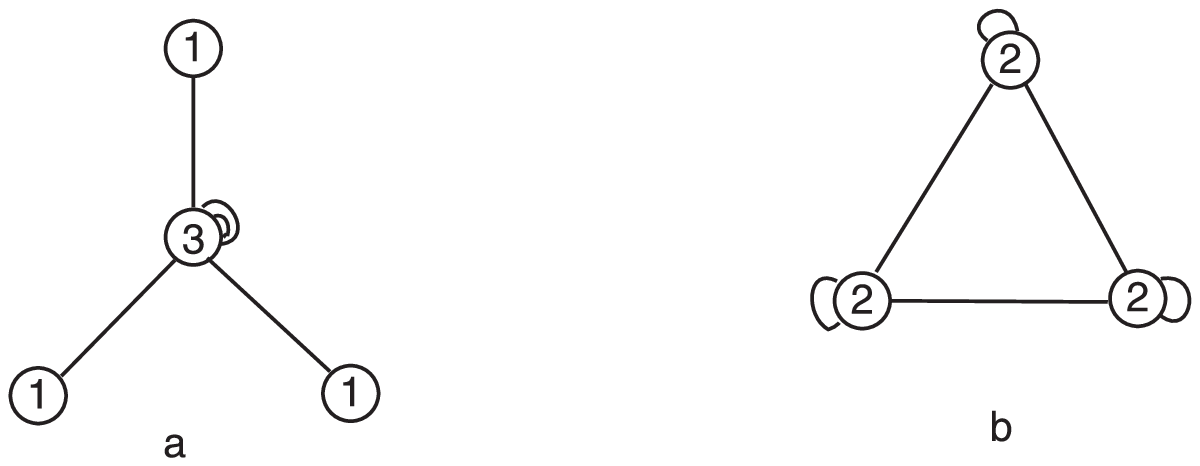}{8}{Building the ${\field E}_6$ quiver: (a) 
without discrete torsion, (b) with discrete torsion.}

The vertical strands of the web indicate another way in which the
$\BE{6}$ and $\BD{2}$ quivers may be conceived. Namely,
\begin{eqnarray}
\BC^2/\BE{6} & \simeq & (\BC^2/\BD{2})/\ZZ_3 \nonumber \\
             & \simeq & (\BC^2/\ZZ_2)/{\field E}_6; \label{eq:e6hat} \\
\BC^2/\BD{2} & \simeq & (\BC^2/\ZZ_4)/\ZZ_2 \nonumber \\ 
             & \simeq & (\BC^2/\ZZ_2)/{\field D}_2. \label{eq:d2hat}
\end{eqnarray}
In the language of Section \ref{sec:prodgrp}, these are examples where
$G_2$ is non-Abelian, and hence, $H^1(G_2, U(1))$ is different from
$G_2$. The orbifold $(\BC^2/\ZZ_2)/{\field E}_6$ can be thought of as an
${\field E}_6$ action on the $\ZZ_2$ lattice.  Each node is identified
with one of the ${\field E}_6$ quivers in Figure \ref{fig:orde6.eps},
and the arrows are such that we recover the $\BE{6}$ quiver from this
construction.  In the case of Eq. (\ref{eq:d2hat}), this prescription is
precisely the reverse of the construction of the ordinary dihedral
quivers discussed in Ref. \cite{amihe}.

\subsubsection{$H^2((\BE{6}\times\ZZ_{2n})/\ZZ_2,U(1)) = \ZZ_3$}

Because both $\BE{6}$ and $\ZZ_{2n}$ contain an element which acts as
\begin{equation}
\begin{pmatrix} -1 & 0 & 0 \\ 0 & -1 & 0 \\ 0 & 0 & 1 \end{pmatrix}
\end{equation}
on the coordinates $(z_1, z_2, z_3)$ of $\BC^3$, we must quotient out by
$\ZZ_2$ just as we did when considering the direct product of the binary
dihedral and cyclic groups.  
Examining the twisted algebra of $(\BE{6}\times\ZZ_{2n})/\ZZ_2$, we find that 
there is a $\ZZ_3$ discrete torsion when $n = 0\ {\rm mod}\ 3$.  The relations
\begin{equation}
e_0 e_1 = \omega_3^a e_1 e_0,\;\;\; 
e_0 e_2 = \omega_3^{-a} e_2 e_0,
\end{equation}
encode the phases, where $e_0$ is the generator of $\ZZ_{2n}$ and $e_1$ and
$e_2$ are generators of $\BE{6}$, whose algebra is given in Eq.
(\ref{eq:E6gens}).  The generators of the product group act on the
coordinates of $\BC^3$ by
\begin{eqnarray}
e_0: && (z_1, z_2, z_3) \rightarrow (\omega_{2n} z_1, \omega_{2n} z_2, \omega_n^{-2} z_3), \nonumber \\
e_1: && (z_1, z_2, z_3) \rightarrow (\Btwo_0(e_1) (z_1, z_2), z_3), \label{eq:act3} \\
e_2: && (z_1, z_2, z_3) \rightarrow (\Btwo_0(e_2) (z_1, z_2), z_3), \nonumber
\end{eqnarray}
where $\Btwo_0(e_i)$ are the generators of the defining representation
of $\BE{6}$ as given in Eqs. (\ref{eq:E6e1}---\ref{eq:E6e2}).  The
superfields $(\phi_1, \phi_2)$ act as $(\omega_{2n}, \Btwo_0)$ while
$\phi_3$ acts as $(\omega_{2n}^{-2}, \Bone_0)$.  The construction of the
$N = 1$ quiver without discrete torsion follows the discussion in
Section \ref{sec:quiver}. The $3n$ one-dimensional and $n$
three-dimensional representations are characterized by even powers of
$\omega_{2n}$ and the $3n$ two-dimensional representations by odd
powers.  In the two cases, the matrix realizations of $e_1$ and $e_2$
are precisely the irreducible representations of the ordinary
tetrahedral group ${\field E}_6$ with trivial and non-trivial $\ZZ_2$
torsion, respectively.  The quiver consists of an interconnected stack
of $n$ copies of the $\BE{6}$ quiver.

Let us briefly apply the same techniques that we have previously
employed to study the quiver of $(\BE{6} \times \ZZ_6)/\ZZ_2$ with the
$\ZZ_3$ discrete torsion turned on.  The torsion acts to produce three
orbits, one cycling the $\Bone$s, another cycling the $\Btwo$s, and the
third leaving the $\Bthree$ fixed.  We therefore expect four $\Bthree$s
and a $\Bsix$ in the final quiver.  Hypermultiplets run between the
$\Bsix$ and each of the $\Bthree$s, and an adjoint sits on each node
formed by joining.  Chiral lines connect the nodes formed by splitting. 
This is illustrated in Figure \ref{fig:e6z3.eps}.
\myfig{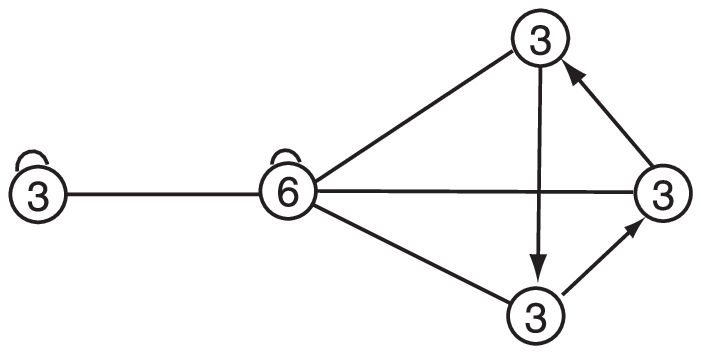}{5}{$(\BE{6} \times \ZZ_6)/\ZZ_2\sim\BE{6}\times\ZZ_3$
 with $\ZZ_3$ discrete torsion.}
The $(\BE{6} \times \ZZ_{6n})/\ZZ_2$ model with discrete
torsion contains $n$ interconnected copies of this quiver.

The story for $(\BE{7}\times\ZZ_{2n})/\ZZ_2$ unfolds along similar lines.
Since $H^2(\BE{7}\times\ZZ_{2n})/\ZZ_2, U(1)) = \ZZ_2$, there is discrete
torsion here as well. 
 
\subsection{$\BC^3/\Delta_{3n^2,6n^2}$}

The group $\Delta_{3n^2}$ is a discrete subgroup of $SU(3)$ given by the
exact sequence
\begin{equation}
0 \rightarrow \BZ_n\times \BZ_n\rightarrow \Delta_{3n^2} \rightarrow \BZ_3\rightarrow 0. \label{eq:del3exact}
\end{equation}
The $\ZZ_3$ acts on the coordinates of $\BC^3$ by permutation and 
the $\ZZ_n$ acts by phases:
\begin{eqnarray}
e_0: && (z_1, z_2, z_3) \rightarrow (z_3, z_1, z_2), \nonumber \\
e_1: && (z_1, z_2, z_3) \rightarrow (\omega_n z_1, \omega_n^{-1} z_2, z_3), \\
e_2: && (z_1, z_2, z_3) \rightarrow (z_1, \omega_n z_2, \omega_n^{-1} z_3). \nonumber
\end{eqnarray}
From this action, we obtain the relations
\begin{eqnarray}
&& e_0^3 = e_1^n = e_2^n = e_1 e_2 e_1^{-1} e_2^{-1} = 1, \\
&& e_0 e_1 e_0^{-1} = e_2,\;\;\; 
   e_0 e_2 e_0^{-1} = e_1^{-1} e_2^{-1}. \label{eq:del3interaction}
\end{eqnarray}

The discrete torsion of $\Delta_{3n^2}$ \cite{amihe} is
\begin{equation}
H^2(\Delta_{3n^2},U(1)) =
\left\{
\begin{matrix}
                        \ZZ_n \times \ZZ_3 && n=0\ {\rm mod}\ 3,\\
                        \ZZ_n && {\rm otherwise}.
\end{matrix}\right.
\end{equation}
The exact sequence (\ref{eq:del3exact}) is a case where $H^2(G_1, U(1))$
is non-trivial.  Thus the discrete torsion arises from two sources: we
may embed a phase in the algebra of $\ZZ_n \times \ZZ_n$, and, when $n =
0\ {\rm mod}\ 3$, there is a second $\BZ_3$ phase 
which may be identified with the `interaction' term of Eq.
(\ref{eq:del3interaction}). Note that $\Delta_{3n^2}$ is {\it not} a
direct product, and as such we should consider where discrete torsion
phases are allowed to enter. The general rule is that they can
be associated with any element of $N$ which commutes with $G/N$, as
$eg=\chi(g) ge$. In the
present case, this is $g=e_1^{n/3} e_2^{2n/3}$, and $\chi$ is a $\ZZ_3$
phase.

We note the special cases $\Delta_{3 \cdot 1^2} \simeq \ZZ_3$ and
$\Delta_{3 \cdot 2^2} \simeq {\field E}_6$ \cite{HanHe1}. Other examples
may be constructed in a similar fashion.

When $n \ne 0\ {\rm mod}\ 3$, the $\Delta_{3n^2}$ quiver is constructed
from the $\ZZ_n \times \ZZ_n$ lattice.  The $\ZZ_3$ acts to produce
$(n^2-1)/3$ orbits among three nodes plus one orbit that leaves a single
node fixed. Thus, in the absence of discrete torsion, the quiver for
$\Delta_{3 \cdot 4^2}$, for example, has five $\Bthree$s and three
$\Bone$s.  The theory is chiral. If we add $\ZZ_n$ discrete torsion, the quiver
is reduced to a $\ZZ_k \times \ZZ_k$ lattice, $k$ depending on the particular
torsion element. Taking the maximal
$\ZZ_4$ torsion on $\Delta_{3 \cdot 4^2}$ for example gives us an
$\BA{2}(4)$ quiver while $\ZZ_2 \subset \ZZ_4$ gives an ${\field
E}_6(2)$ quiver.\footnote{Here the notation $G(n)$ refers to the quiver
obtained from the quiver of $G$ by mulitplying the rank of each node by
$n$ (Morita equivalence).}

If $n = 0\ {\rm mod}\ 3$, the $\ZZ_3$ acts on the $\ZZ_n \times \ZZ_n$
lattice to give nine one-dimensional and $n^2/3 - 1$ three-dimensional
irreducible representations \cite{Muto}. If we consider the maximal
discrete torsion, then we get the same result as before, namely, the
quiver diagram of the $\BZ_3$ orbifold where the nodes are now of rank
$n$ ({\em i.e.} $\BA{2}(n)$). We are interested in studying how the
second $\ZZ_3$ torsion acts on the $\ZZ_n\times\ZZ_n$ quiver.

Let us now specialize to the case $n = 3$.  Without discrete torsion,
there are three fixed nodes and two orbits of order three under the
action of $e_0$ on the $\ZZ_3 \times \ZZ_3$ lattice.  Thus, we obtain
the quiver of $\Delta_{27}$ in Figure \ref{fig:Delta27.eps}.
\myfig{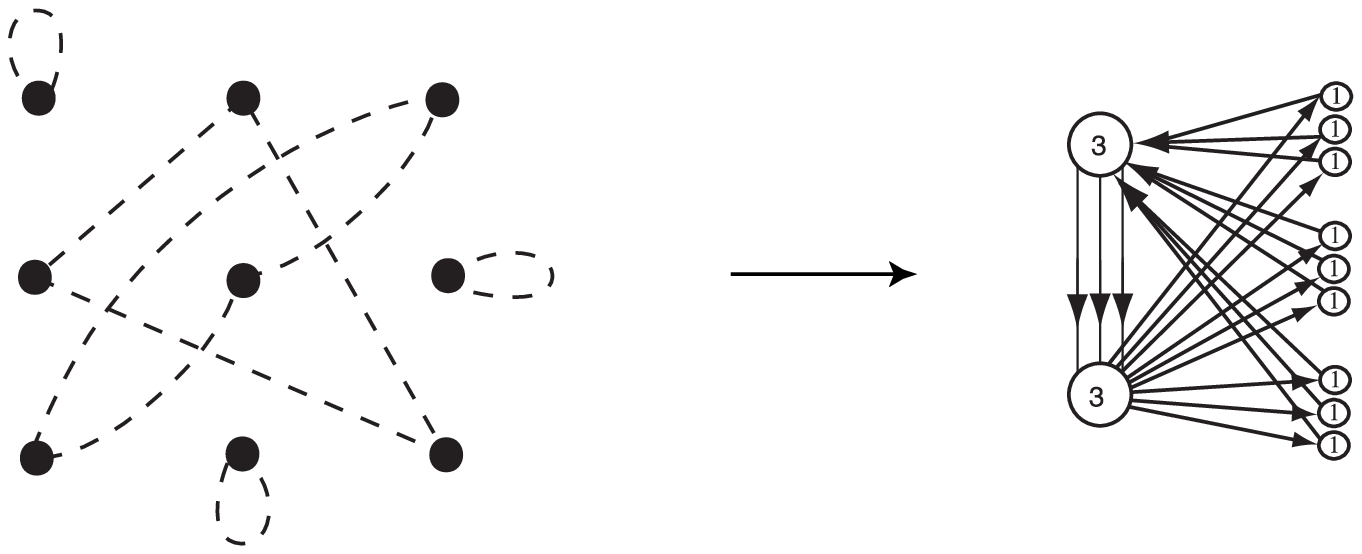}{8}{A $\BZ_3$ automorphism of the $\ZZ_3 \times
\ZZ_3$ lattice gives the quiver of $\Delta_{27}$ (no discrete torsion).}

If we admit the discrete torsion of the second $\ZZ_3$, then we change
the $\ZZ_3$ orbits. We find that this produces three 3-node orbits, 
which consist of one
of the 1-node orbits together with a node from each of the
3-node orbits from before.  Thus, the resulting quiver is also
$\BA{2}(3)$.  We get the {\em same} quiver diagram for any choice
of discrete torsion phases.  However, the superpotentials of two
such quivers must differ because string theory is sensitive to the
choices we have made.

The group $\Delta_{6n^2}$, which is a subgroup of $SU(3)$ for even $n$
\cite{Greene}, fits into the exact sequence
\begin{equation}
0 \rightarrow \Delta_{3n^2} \rightarrow \Delta_{6n^2} \rightarrow \ZZ_2 \rightarrow 0.
\end{equation}
Hence, its quiver may be constructed from the $\Delta_{3n^2}$ quiver using
the techniques we have discussed.  The authors of Ref. \cite{amihe} calculate
that $H^2(\Delta_{6n^2}, U(1)) = \ZZ_2$.

\section{Dualities} \label{sec:dualities}

We have seen that different orbifolds may have the same quiver
diagram. Thus, there are different orbifold points in the moduli space of
couplings of the corresponding gauge theory. Then from the AdS/CFT
correspondence \cite{M, GKP, W}, we obtain the result that different
orbifolds are on the same moduli space.

A non-trivial example is provided by the $\BA{2}$ quiver,
which is shown in Figure \ref{fig:delta3n.eps}.
\myfig{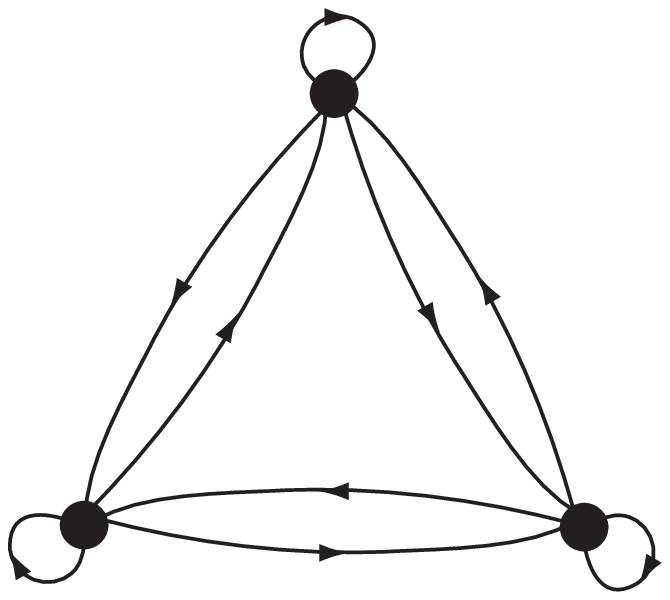}{2}{The $\BA{2}$ quiver.}
This is the quiver for the $N=2$ model with superpotential
\begin{equation}\label{eq:zeromarg}
W_1=\tr [\phi_1,\phi_2]\phi_3.
\end{equation}
Here, $\phi_1$ is a $3m\times 3m$ matrix
representing the adjoint fields, while $\phi_2$ and $\phi_3$ contain the
bifundamental fields.

Alternatively, we may obtain the same quiver from a $\ZZ_{3n}\times\ZZ_n$ orbifold.
Here we choose maximal discrete torsion to obtain
the $U(m)^3$ model. The superpotential here may be written
\begin{equation}\label{eq:firstmarg}
W_1=\tr [\phi_1,\phi_2]\phi_3+\left( \frac{1-q}{1+q}\right) \tr \{\phi_1,\phi_2\}\phi_3,
\end{equation}
where $q$ is a $3n$-th root of unity. 

From the above discussions, we know that $\Delta_{3n^2}$ may be thought
of in terms of the exact sequence $0\to
\ZZ_n\times\ZZ_n\to\Delta_{3n^2}\to\ZZ_3\to 0$. As a result, the theory
is constructed by a $\ZZ_3$ projection of the $\ZZ_n\times\ZZ_n$
orbifold. The $\ZZ_3$ permutes the three fields $\phi_{1,2,3}$. This gives rise
again to the $\BA{2}$ quiver. We may
define fields $\chi_{1,2,3}$ which transform by rephasing; in terms of
these fields, which have the same interpretation as the fields of
the $\ZZ_{3n}\times\ZZ_n$ orbifold, we find a superpotential
\begin{equation}\label{eq:secmarg}
W_2=\tr[\chi_1,\chi_2]\chi_3+\left(\frac{1-q^3}{1+q^3}\right) \left[\frac{i}{\sqrt{3}}\
\tr \{\chi_1,\chi_2\}\chi_3+\frac{1}{3}\tr \left( \chi_1^3+\chi_2^3+\chi_3^3\right)\right].
\end{equation}
The superpotential (\ref{eq:firstmarg}) is a marginal perturbation of (\ref{eq:zeromarg}),
reminiscent of Refs. \cite{BJL1,BJL2}. The superpotential (\ref{eq:secmarg}) is more interesting:
it contains {\it both} of the marginal perturbations of Ref. \cite{ls}.

\section{Conclusion} \label{sec:concl}

In this paper we have presented a technique which permits calculations of quiver
diagrams for certain orbifold singularities very efficiently. Presumably 
these techniques can also be applied to orientifolds with discrete torsion
\cite{Klein1,Klein2} and to other singularities such as orbifolds of
conifolds \cite{Dasg}.

Quiver diagrams inherit the quantum symmetry of an orbifold, so one can also
orbifold a quiver by a subgroup of its quantum symmetry. Although we did not show
this explicitly in the paper, our techniques applied in this case give
the correct quiver diagram of the (partially)  unorbifolded orbifold.

With some other results in the literature \cite{HanHe1, HanHe2, FHH-1}
these techniques should provide a complementary set of tools to 
study D-brane field theories. We note also that after this work had been
completed, we became aware of the paper \cite{newHan} which
addresses similar issues using different techniques.

\bigskip \noindent {\bf Acknowledgments:} We thank Bo Feng, Amihay Hanany,
Yang-Hui He and Cumrun Vafa for discussions. Our work is supported in part by the U.S.
Department of Energy, grant DE-FG02-91ER40677, and an Outstanding Junior
Investigator Award. VJ is supported by a GAANN fellowship from the Department
of Education, grant 1533616.

\section{Appendix $\field E$}\label{sec:AppE}

\noindent The binary tetrahedral group $\BE{6}$ is defined by the relations
\cite{Cox-1}
\begin{equation}
e_1^3 = e_2^3 = (e_2 e_1)^2 \label{eq:E6gens}.
\end{equation}
The one-dimensional irreducible representations are
\begin{equation}
{\bf 1}_a:\; (e_1, e_2) = (\omega_3^a, \omega_3^{-a})
\end{equation}
for $a = 0, 1, 2$.  In terms of the quaternions, one can write \cite{Cox-2}
the generators of a two-dimensional irreducible representation (the defining
representation, $\Btwo_0$) as
\begin{eqnarray}
e_1 & = & \frac{1}{2}(1 + i + j + k), \label{eq:E6e1} \\
e_2 & = & \frac{1}{2}(1 + i + j - k). \label{eq:E6e2} 
\end{eqnarray}
The other $\Btwo$s are defined as $\Btwo_a = \Btwo_0 \otimes \Bone_a.$ There
is also a three-dimensional irreducible representation:  $\Btwo_0 \otimes
\Btwo_a = \Bthree \oplus \Bone_a$.  The 24 elements of $\BE{6}$ are
\begin{equation}
\left\{\pm 1, \pm i, \pm j, \pm k, \frac{1}{2}(\pm 1 \pm i \pm j \pm k)\right\},
\end{equation}
and the commutator subgroup $\BE{6}' \simeq \BD{2}$ consists of the first
eight elements in this list\footnote{This fact, coupled with the exact
sequence Eq. (\ref{eq:Reid-6}), immediately gives the discrete torsion of
$\BE{6} \times \ZZ_n$ quoted in Ref. \cite{amihe}.  However, when $n$ is even
neither this group nor $\BE{7} \times \ZZ_n$ is in $SU(3)$.}.

The ordinary tetrahedral group ${\field E}_6$ is a discrete subgroup of  
$SU(2)/\ZZ_2 \simeq SO(3)$.  We now require that 
\begin{equation}
e_1^3 = e_2^3 = (e_2 e_1)^2 = 1.
\end{equation}
Only the one- and three-dimensional irreducible representations of $\BE{6}$
satisfy these relations.  The two-dimensional irreducible representations
correspond to including a non-trivial $\ZZ_2$ torsion in the previous
relations.  That is to say, they satisfy
\begin{equation}
e_1^3 = e_2^3 = (e_2 e_1)^2 = -1.
\end{equation}

The representation theory of the binary octahedral group $\BE{7}$ proceeds
along similar lines.  Here, the presentation \cite{Cox-1} is
\begin{equation}
e_1^4 = e_2^3 = (e_2 e_1)^2. \label{eq:E7gens}
\end{equation}
The one-dimensional irreducible representations are then 
\begin{equation}
{\bf 1}_0:\; (e_1, e_2) = (+1, +1),\;\;\;
{\bf 1}_1:\; (e_1, e_2) = (-1, +1),
\end{equation}
and the defining representation in terms of quaternions 
\cite{Cox-2} is
\begin{eqnarray}
e_1 & = & \frac{1}{\sqrt{2}}(1 + i),\\
e_2 & = & \frac{1}{2}(1 + i + j + k).
\end{eqnarray}
Tensor products yield two other $\Btwo$s, two $\Bthree$s, and a $\Bfour$.
The 48 elements of $\BE{7}$ are
\begin{eqnarray}
&& \left\{\pm 1, \pm i, \pm j, \pm k, \frac{1}{2}(\pm 1 \pm i \pm j \pm k),
\frac{1}{\sqrt{2}}(\pm 1 \pm i), \frac{1}{\sqrt{2}}(\pm 1 \pm j),\right. \nonumber\\
&& \left.\; \frac{1}{\sqrt{2}}(\pm 1 \pm k), \frac{1}{\sqrt{2}}(\pm i \pm j), 
\frac{1}{\sqrt{2}}(\pm i \pm k), \frac{1}{\sqrt{2}}(\pm j \pm k)\right\},
\end{eqnarray}
and the commutator subgroup $\BE{7}' \simeq \BE{6}$.  The two $\Bone$s, one
of the $\Btwo$s (not the defining representation), and the two $\Bthree$s
satisfy the relations of the ordinary octahedral group:
\begin{equation}
e_1^4 = e_2^3 = (e_2 e_1)^2 = 1.
\end{equation}
The other irreducible representations have
\begin{equation}
e_1^4 = e_2^3 = (e_2 e_1)^2 = -1.
\end{equation}

By the McKay Correspondence \cite{McKay}, the quivers of the binary
polyhedral groups correspond to affine Dynkin diagrams of the exceptional Lie
groups.

\pagebreak

\providecommand{\href}[2]{#2}\begingroup\raggedright\endgroup

\begin{thebibliography}{10}

\bibitem{KKV}
S.~Katz, A.~Klemm, and C.~Vafa, ``{{\it Geometric engineering of quantum field
  theories,}}'' {\em Nucl. Phys.} {\bf B497} (1997) 173--195,
  \href{http://xxx.lanl.gov/abs/hep-th/9609239}{{hep-th/9609239}}.

\bibitem{KV}
S.~Katz and C.~Vafa, ``{{\it Geometric engineering of N = 1 quantum field
  theories,}}'' {\em Nucl. Phys.} {\bf B497} (1997) 196--204,
  \href{http://xxx.lanl.gov/abs/hep-th/9611090}{{hep-th/9611090}}.

\bibitem{BJPSV}
M.~Bershadsky, A.~Johansen, T.~Pantev, V.~Sadov, and C.~Vafa, ``{{\it F-theory,
  geometric engineering and N = 1 dualities,}}'' {\em Nucl. Phys.} {\bf B505}
  (1997) 153--164,
  \href{http://xxx.lanl.gov/abs/hep-th/9612052}{{hep-th/9612052}}.

\bibitem{KMV}
S.~Katz, P.~Mayr, and C.~Vafa, ``{{\it Mirror symmetry and exact solution of 4D
  N = 2 gauge theories. I,}}'' {\em Adv. Theor. Math. Phys.} {\bf 1} (1998)
  53--114, \href{http://xxx.lanl.gov/abs/hep-th/9706110}{{hep-th/9706110}}.

\bibitem{BDL}
M.~Berkooz, M.~R. Douglas, and R.~G. Leigh, ``{{\it Branes intersecting at
  angles,}}'' {\em Nucl. Phys.} {\bf B480} (1996) 265--278,
  \href{http://xxx.lanl.gov/abs/hep-th/9606139}{{hep-th/9606139}}.


\bibitem{bgkl}
R. Blumenhagen, L. Goerlich, B. Kors and D. Lust,
 ``{{\it Noncommutative compactifications of type I strings on tori 
 with  magnetic background flux,}}'' {\em JHEP} {\bf 10} (2000) 6,
  \href{http://xxx.lanl.gov/abs/hep-th/0007024}{{hep-th/0007024}}.


\bibitem{Ur1}
G.~Aldazabal, S.~Franco, L.~E. Ibanez, R.~Rabadan, and A.~M. Uranga, ``{{\it D
  = 4 chiral string compactifications from intersecting branes,}}''
  \href{http://xxx.lanl.gov/abs/hep-th/0011073}{{hep-th/0011073}}.

\bibitem{Ur2}
G.~Aldazabal, S.~Franco, L.~E. Ibanez, R.~Rabadan, and A.~M. Uranga, ``{{\it
  Intersecting brane worlds,}}''
  \href{http://xxx.lanl.gov/abs/hep-ph/0011132}{{hep-ph/0011132}}.

\bibitem{DM}
M.~R. Douglas and G.~Moore, ``{{\it D-branes, Quivers, and ALE Instantons,}}''
  \href{http://xxx.lanl.gov/abs/hep-th/9603167}{{hep-th/9603167}}.

\bibitem{V}
C.~Vafa, ``{{\it Modular Invariance And Discrete Torsion On Orbifolds,}}'' {\em
  Nucl. Phys.} {\bf B273} (1986) 592.

\bibitem{D}
M.~R. Douglas, ``{{\it D-branes and discrete torsion,}}''
  \href{http://xxx.lanl.gov/abs/hep-th/9807235}{{hep-th/9807235}}.

\bibitem{DF}
M.~R. Douglas and B.~Fiol, ``{{\it D-branes and discrete torsion. II,}}''
  \href{http://xxx.lanl.gov/abs/hep-th/9903031}{{hep-th/9903031}}.

\bibitem{BL}
D.~Berenstein and R.~G. Leigh, ``{{\it Discrete torsion, AdS/CFT and
  duality,}}'' {\em JHEP} {\bf 01} (2000) 038,
  \href{http://xxx.lanl.gov/abs/hep-th/0001055}{{hep-th/0001055}}.

\bibitem{BJL1}
D.~Berenstein, V.~Jejjala, and R.~G. Leigh, ``{{\it Marginal and relevant
  deformations of N = 4 field theories and non-commutative moduli spaces of
  vacua,}}'' {\em Nucl. Phys.} {\bf B589} (2000) 196--248,
  \href{http://xxx.lanl.gov/abs/hep-th/0005087}{{hep-th/0005087}}.

\bibitem{BJL2}
D.~Berenstein, V.~Jejjala, and R.~G. Leigh, ``{{\it Noncommutative moduli
  spaces, dielectric tori and T duality,}}''
  \href{http://xxx.lanl.gov/abs/hep-th/0006168}{{hep-th/0006168}}.

\bibitem{G}
J.~Gomis, ``{{\it D-branes on orbifolds with discrete torsion and topological
  obstruction,}}'' {\em JHEP} {\bf 05} (2000) 006,
  \href{http://xxx.lanl.gov/abs/hep-th/0001200}{{hep-th/0001200}}.

\bibitem{Asp1}
P.~S. Aspinwall and M.~R. Plesser, ``{{\it D-branes, discrete torsion and the
  McKay correspondence,}}''
  \href{http://xxx.lanl.gov/abs/hep-th/0009042}{{hep-th/0009042}}.

\bibitem{W2}
E.~Witten, ``{{\it New *gauge* theories in six dimensions,}}'' {\em JHEP} {\bf
  01} (1998) 001,
  \href{http://xxx.lanl.gov/abs/hep-th/9710065}{{hep-th/9710065}}.

\bibitem{BCD}
D.~Berenstein, R.~Corrado, and J.~Distler, ``{{\it Aspects of ALE matrix models
  and twisted matrix strings,}}'' {\em Phys. Rev.} {\bf D58} (1998) 026005,
  \href{http://xxx.lanl.gov/abs/hep-th/9712049}{{hep-th/9712049}}.

\bibitem{amihe}
B.~Feng, A.~Hanany, Y.-H. He, and N.~Prezas, ``{{\it Discrete torsion,
  non-Abelian orbifolds and the Schur multiplier,}}''
  \href{http://xxx.lanl.gov/abs/hep-th/0010023}{{hep-th/0010023}}.

\bibitem{Cox-1}
H.~Coxeter and W.~Moser, {\em Generators and Relations for Discrete Groups}.
\newblock Springer-Verlag, Berlin, 1972.

\bibitem{Reid}
M.~Reid, ``{{\it Young person's guide to canonical singularities,}}'' {\em
  Proceedings of Symposia in Pure Mathematics.} {\bf 46} (1985) 345.

\bibitem{Kar1}
G.~Karpilovsky, {\em Projective Representations of Finite Groups}.
\newblock Marcel Dekker, Inc., New York, 1985.

\bibitem{Adem}
A.~Adem and R.~J. Milgram, {\em Cohomology of finite groups}.
\newblock Springer-Verlag, Berlin, 1994.

\bibitem{VW}
C.~Vafa and E.~Witten, ``{{\it On orbifolds with discrete torsion,}}'' {\em J.
  Geom. Phys.} {\bf 15} (1995) 189,
  \href{http://xxx.lanl.gov/abs/hep-th/9409188}{{hep-th/9409188}}.

\bibitem{MukRay}
S.~Mukhopadhyay and K.~Ray, ``{{\it D-branes on fourfolds with discrete
  torsion,}}'' {\em Nucl. Phys.} {\bf B576} (2000) 152--176,
  \href{http://xxx.lanl.gov/abs/hep-th/9909107}{{hep-th/9909107}}.

\bibitem{FHH-1}
B.~Feng, A.~Hanany, and Y.-H. He, ``{{\it The Z(k) x D(k') brane box model,}}''
  {\em JHEP} {\bf 09} (1999) 011,
  \href{http://xxx.lanl.gov/abs/hep-th/9906031}{{hep-th/9906031}}.

\bibitem{FHH-2}
B.~Feng, A.~Hanany, and Y.-H. He, ``{{\it Z-D brane box models and non-chiral
  dihedral quivers,}}''
  \href{http://xxx.lanl.gov/abs/hep-th/9909125}{{hep-th/9909125}}.

\bibitem{Kar2}
G.~Karpilovsky, {\em The Schur Multiplier}.
\newblock Clarendon Press, Oxford, 1987.

\bibitem{HanHe1}
A.~Hanany and Y.-H. He, ``{{\it Non-Abelian finite gauge theories,}}'' {\em
  JHEP} {\bf 02} (1999) 013,
  \href{http://xxx.lanl.gov/abs/hep-th/9811183}{{hep-th/9811183}}.

\bibitem{Muto}
T.~Muto, ``{{\it D-branes on three-dimensional nonabelian orbifolds,}}'' {\em
  JHEP} {\bf 02} (1999) 008,
  \href{http://xxx.lanl.gov/abs/hep-th/9811258}{{hep-th/9811258}}.

\bibitem{Greene}
B.~R. Greene, C.~I. Lazaroiu, and M.~Raugas, ``{{\it D-branes on nonabelian
  threefold quotient singularities,}}'' {\em Nucl. Phys.} {\bf B553} (1999)
  711--749, \href{http://xxx.lanl.gov/abs/hep-th/9811201}{{hep-th/9811201}}.

\bibitem{M}
J.~Maldacena, ``{{\it The large-N limit of superconformal field theories and
  supergravity,}}'' {\em Adv. Theor. Math. Phys.} {\bf 2} (1998) 231--252,
  \href{http://xxx.lanl.gov/abs/hep-th/9711200}{{hep-th/9711200}}.

\bibitem{GKP}
S.~S. Gubser, I.~R. Klebanov, and A.~M. Polyakov, ``{{\it Gauge theory
  correlators from non-critical string theory,}}'' {\em Phys. Lett.} {\bf B428}
  (1998) 105, \href{http://xxx.lanl.gov/abs/hep-th/9802109}{{hep-th/9802109}}.

\bibitem{W}
E.~Witten, ``{{\it Anti-de Sitter space and holography,}}'' {\em Adv. Theor.
  Math. Phys.} {\bf 2} (1998) 253--291,
  \href{http://xxx.lanl.gov/abs/hep-th/9802150}{{hep-th/9802150}}.

\bibitem{ls}
R.~G. Leigh and M.~J. Strassler, ``{{\it Exactly marginal operators and duality
  in four-dimensional N=1 supersymmetric gauge theory,}}'' {\em Nucl. Phys.}
  {\bf B447} (1995) 95,
  \href{http://xxx.lanl.gov/abs/hep-th/9503121}{{hep-th/9503121}}.

\bibitem{Klein1}
M.~Klein and R.~Rabadan, ``{{\it Orientifolds with discrete torsion,}}'' {\em
  JHEP} {\bf 07} (2000) 040,
  \href{http://xxx.lanl.gov/abs/hep-th/0002103}{{hep-th/0002103}}.

\bibitem{Klein2}
M.~Klein and R.~Rabadan, ``{{\it Z(N) x Z(M) orientifolds with and without
  discrete torsion,}}'' {\em JHEP} {\bf 10} (2000) 049,
  \href{http://xxx.lanl.gov/abs/hep-th/0008173}{{hep-th/0008173}}.

\bibitem{Dasg}
K.~Dasgupta, S.~Hyun, K.~Oh, and R.~Tatar, ``{{\it Conifolds with discrete
  torsion and noncommutativity,}}'' {\em JHEP} {\bf 09} (2000) 043,
  \href{http://xxx.lanl.gov/abs/hep-th/0008091}{{hep-th/0008091}}.

\bibitem{HanHe2}
A.~Hanany and Y.-H. He, ``{{\it A monograph on the classification of the
  discrete subgroups of SU(4),}}''
  \href{http://xxx.lanl.gov/abs/hep-th/9905212}{{hep-th/9905212}}.

\bibitem{newHan}
B.~Feng, A.~Hanany, Y.-H. He, and N.~Prezas, ``{{\it Discrete torsion, covering
  groups and quiver diagrams,}}''
  \href{http://xxx.lanl.gov/abs/hep-th/0011192}{{hep-th/0011192}}.

\bibitem{Cox-2}
H.~Coxeter, ``{{\it The binary polyhedral groups and other generalizations of
  the quaternion group,}}'' {\em Duke Math J.} {\bf 7} (1940) 367.

\bibitem{McKay}
J.~McKay, ``{{\it Graphs, Singularities, and Finite Groups,}}'' {\em
  Proceedings of Symposia in Pure Mathematics.} {\bf 37} (1980) 183.

\end{thebibliography}

\end{document}